# Lithium transport through Lithium-ion battery cathode coatings

Shenzhen Xu(‡,*), Ryan M. Jacobs(‡,*), Ha M. Nguyen(†,*), Shiqiang Hao(§), Mahesh Mahanthappa(§§,‡), Chris Wolverton(§), Dane Morgan(†,‡)

†Department of Materials Science and Engineering, University of Wisconsin, Madison, Wisconsin
‡Materials Science Program, University of Wisconsin, Madison, Wisconsin
§§Department of Chemistry, University of Wisconsin, Madison, Wisconsin
§Department of Materials Science and Engineering, Northwestern University, Evanston, Illinois
* These authors contributed equally to this work

**ABSTRACT:** The surface coating of cathodes using insulator films has proven to be a promising method for high-voltage cathode stabilization in Li-ion batteries, but there is still substantial uncertainty about how these films function. More specifically, there is limited knowledge of lithium solubility and transport through the films, which is important for coating design and development. This study uses first-principles calculations based on Density Functional Theory to examine the diffusivity of interstitial lithium in the crystals of α-AlF$_3$, α-Al$_2$O$_3$, m-ZrO$_2$, c-MgO, and α-quartz SiO$_2$, which provide benchmark cases for further understanding of insulator coatings in general. In addition, we propose an Ohmic electrolyte model to predict resistivities and overpotential contributions under battery operating conditions. For the crystalline materials considered we predict that Li$^+$ diffuses quite slowly, with a migration barrier larger than 0.9 eV in all crystalline materials except α-quartz SiO$_2$, which is predicted to have a migration barrier of 0.276 eV along <001>. These results suggest that the stable crystalline forms of these insulator materials, except for oriented α-quartz SiO$_2$, are not practical for conformal cathode coatings. Amorphous Al$_2$O$_3$ and AlF$_3$ have higher Li$^+$ diffusivities than their crystalline counterparts. Our predicted amorphous Al$_2$O$_3$ resistivity (1789 MΩm) is close to the top of the range of the fitted resistivities extracted from previous experiments on nominal Al$_2$O$_3$ coatings (7.8 to 913 MΩm) while our predicted amorphous AlF$_3$ resistivity (114 MΩm) is very close to the middle of the range. These comparisons support our framework for modeling and understanding the impact on overpotential of conformal coatings in terms of their fundamental thermodynamic and kinetic properties, and support that these materials can provide practical conformal coatings in their amorphous form.

## I. Introduction

Surface modification of the cathode by artificial coating is an effective strategy to stabilize Li-ion batteries (LIBs) operating at high voltages.[1-10] Nevertheless, the coating functionalities and the stabilizing mechanisms are still not fully understood and currently a subject of intensive research in the development of next-generation LIBs. Several roles for the coating have been proposed to account for its positive impacts on the cathode performance, including: (i) electrical conduction medium that facilitates electron transport between cathode active particles,[11] (ii) modifier of cathode surface chemistry that changes chemical properties of the cathode surface to improve stability and performance,[12] (iii) HF scavenger that locally reduces the acidity of the electrolyte near the cathode surface, thereby reducing electrolyte degradation,[13-15] and (iv) physical protection barrier that suppresses electrolyte oxidation and cathode corrosion.[8,16-20] In general, role (iii) and, particularly, role (iv) are the most widely claimed origins of enhanced performance. The coatings must also allow adequate electrical and lithium transport if they cover all or even most of the surface of each particle, and the extent of coverage required has not been established. As it is still not clear all of the roles a coating might or does play in improving performance, developing an optimal coating is particularly challenging. Furthermore, some coating properties may enhance one aspect of performance while hindering another. For example, higher electrical conductivity may be beneficial if particles have coating between themselves and the conducting matrix in the electrode,[21] but reduce coating effectiveness if the coating is primarily protecting against electrolyte oxidation by the cathode.[5] In practice, many coating materials with varying properties and conformity have been found to improve cathode performance as measured by both increased capacity and capacity retention.[2,3,5,6,19]

For the purposes of this work, we will assume that the cathode coating primarily works through creating a barrier layer against electron and perhaps ion transport (cathode dissolution), in effect reducing electrolyte oxidation and cathode corrosion, respectively. Therefore, extensive coverage of the cathode/electrolyte interface is valuable for a coating to be effective. Lithium transport between the electrolyte and cathode will take place either through the limited uncoated regions or through the coating itself, either through its bulk or short-circuit paths like grain boundaries, pinholes, etc. Assuming one would like to maximize the coated regions, the ability of the coating to transport lithium is likely to play an important role.[22] As an example of where Li transport may be limiting, a conformal ultrathin $Al_2O_3$ film coated on a $LiCoO_2$ cathode was shown to enhance performance when very thin but reduce the performance when the coating became thicker than about 0.5 nm (i.e., about 4 Atomic Layer Deposition (ALD) cycles of a typical growth rate of 0.11-0.12 nm/cycle).[19] Therefore, understanding lithium ion transport through insulator coating films as a function of coating thickness, atomic structure, coating/cathode interfacial heterostructure, and defect chemistry[23-26] is of importance in the development of cathode coatings.

In this paper, we use first-principles calculations to investigate lithium ion transport through a number of idealized inorganic insulator materials that have been explored for cathode coating,[2,4] with a focus on $AlF_3$,[27-30] $Al_2O_3$,[17,18] $MgO$,[31,32] $SiO_2$,[33-35] and $ZrO_2$.[16,36] Coating films are frequently found to be at least partially or fully amorphous,[2,3,17,18] although their structures will typically depend on the exact synthesis methods and conditions, some of which can result in coating films of nanocrystallite morphologies.[16,36] This work has focused primarily on crystalline films as these are the simplest to study with atomistic modeling methods and provide a benchmark case for considering coating performance with more complex nanostructures. We also build on previous published diffusion calculations to consider the performance of select amorphous (prefixed with "am") films, including am-$Al_2O_3$, am-$Li_{3.5}Al_2O_3$, and am-$AlF_3$.[37,38] The behavior of Li diffusion and the resulting coating resistivities through crystalline and select amorphous materials are compared to elucidate the role atomic structure may play in realizing an effective cathode coating.

The paper is organized as follows. Sec. II gives the computational details and discusses the relevant models for Li transport through coating materials. Sec. IIIA discusses the Li migration, focusing on just the crystalline phases. Sec. IIIB discusses the analysis of the amorphous systems, which is done separately from the crystalline systems as their treatment involves a number of different approaches than used for the crystalline systems. Discussion of the implications of the results and analysis with an Ohmic electrolyte model is given in Sec IIIC and conclusions are given in Sec. IV.

## II. Computational methods

### A. Density functional theory calculations

We use Density Functional Theory (DFT) methods as implemented in the Vienna Ab Initio Simulation Package (VASP) to calculate lithium defect formation and migration energies in a series of oxide coatings.[39-41] VASP calculations are performed with the projector-augmented wave (PAW) method[42,43] using the Perdew-Wang (PW91) version of the Generalized Gradient Approximation (GGA) exchange-correlation potentials[44] and a cutoff energy for the planewave basis functions of 600 eV. The pseudopotentials and valence electron configurations of the atoms used are Li (Li_sv and $2s^22p^1$), Al (Al and $2s^22p^1$), Mg (Mg and $2s^22p^0$), Si ($2s^22p^2$), Zr (Zr_sv and $4s^24p^65s^24d^2$), F (F and $2s^22p^5$), and O (O and $2s^22p^4$). A 5 x 5 x 5 Monkhorst-Pack k-point mesh is used for sampling the Brillouin zone of the reciprocal space for all supercells. Supercells are 2 x 2 x 2 primitive cells of the respective coating structures, except for MgO where a 3 x 3 x 3 supercell is used (each supercell contains about 60 to 100 atoms in total). The atomic positions are fully relaxed to minimize the total energy until it converges within an accuracy of better than 1 meV per cell. Plane wave energy cutoff and k-point mesh density were separately tested and were also converged to give a total energy within 1 meV per cell. For these calculations, we consider each coating material in its most stable crystalline form, as shown in Fig. 1.

**B.    Li migration in crystalline coatings**

In order to search for possible migration pathways and obtain the associated migration barriers, $E_m$, for the lithium ion diffusion, we employ the climbing image nudged elastic band (CI-NEB)[45] method as implemented in VASP for GGA calculations. The images of the CI-NEB are relaxed internally until the maximum residual force is less than 0.01 eV/Å but no volume or cell parameter relaxations are performed on the images during optimization.

We estimate the effective diffusivity from the simple Arrhenius form[46]

$$D = D_0 exp(-\beta E_m) \approx a^2 \nu exp(-\beta E_m) \qquad (1)$$

Here, $a$ is the hop distance and $\nu$ is the phonon frequency. The second equality is approximately true for a diffusing dilute interstitial, where we have set correlation and geometric factors to be unity, an approximation that is not expected to alter values by an amount significant for this study. We note that this formula yields a single diffusion coefficient and is therefore only rigorously applicable to isotropic diffusion. We predict isotropic diffusion for c-MgO and α-AlF$_3$, but expect anisotropic diffusion for α-Al$_2$O$_3$, m-ZrO$_2$, and α-quartz SiO$_2$. Consistent with the limited knowledge we have of atomic structure, nano- and micro-structure, defect properties, and operating mechanisms of these coatings, our goal in this work is to provide semi-quantitative understanding of Li in these coating materials. We therefore focus on obtaining an approximate upper bound of the diffusivity along any direction, rather than the full anisotropic diffusion tensor. In this spirit, we simply apply Eq. (1) using the lowest barrier for Li hopping that allows for diffusive motion along any direction. As an exception to this approach, we quote two values for α-quartz SiO$_2$, as this material shows such an anomalously fast diffusion along one direction it is helpful to know what the next limiting barrier is likely to be if full 3D diffusion is required. Also in the spirit of a semi-quantitative model we take $a = 5$ Å, $\nu = 10^{13}$ Hz (yielding $D_0 = 2.5 \times 10^{-2}$ cm$^2$/s) for all the crystalline diffusion calculations, as these will likely vary by less than a factor of five, and the dominant factor governing the Li transport is $E_m$. For the amorphous diffusion calculations, discussed further in Sec. IIIC, we use $D_0$ and $E_m$ values from Hao and Wolverton[37] and Jung, et. al..[38]

**C.    Models of Li transport in conformal coatings**

The coatings considered in this work are all nominally insulators, typically with significant band gaps of a few eV or more. Therefore, they might reasonably be expected to be strong insulators. However, we are particularly focused on thin conformal coatings applied by ALD, where the defect chemistry and impurity content could potentially lead to significant trapped charges and some electronic conductivity. Therefore, it is not obvious how one should model the nature of the coating electronic and ionic conduction, and different choices can lead to different models for assessing the impact of Li diffusivity on potential drop across the coating. Here, we consider three distinct possibilities for Li transport mechanisms though the conformal coating: (1) the coating is electronically conducting, no electric field exists in the coating, and Li transport in the form of Li$^0$ is driven only by its own concentration gradient. We will refer to this transport mechanism as the "electron-conducting model". (2) The coating is electronically insulating, Li$^+$ is the only mobile species and its transport is driven by an electric field. No negative compensating charge exists in the coating, therefore a space charge develops the coating. We call this transport mechanism the "space-charge model". (3) The coating is electronically insulating, Li$^+$ is the only mobile species and its transport is driven by an electric field. However, a negative compensating charge exists in the coating and balances the charge of Li$^+$. For simplicity we assume these negative charges are immobile. This model treats the coating like an electrolyte, and will be referred to as the "electrolyte model" or "Ohmic electrolyte model". Next, we analyze the characteristics of each of these proposed models to ascertain which model is the most physically correct for conformal coatings in lithium ion batteries. In this analysis we will focus on am-Al$_2$O$_3$ as a representative example for assessing the models as it is the most widely studied ALD deposited conformal coating to date.

Significant problems with the electron-conducting model can be shown by a simple estimation of flux and comparison to experimental values. We estimate the Li$^0$ flux in a simple one-dimensional steady state case where it can be determined from $j = -D\frac{dC(x)}{dx}$, where $j$ is the flux density, $D$ is the Li$^0$ diffusivity in the coating, and $C(x)$ is the Li$^0$ concentration as a function of spatial coordinate $x$. In a charging/discharging process, if the system is assumed to be in a steady state, the $Li^0$ flux inside the coating should give a current density $J = ej$, where $e$ is the unit charge of one $Li^+$. Under steady state conditions, the flux $j$ is a constant with respect to the coordinate $x$, therefore the concentration gradient is a constant value through the thickness of the coating. We can make an estimation of the maximum Li flux density under steady state conditions. Let $C_{max}$ denote the total intercalation site concentration in a coating material. The upper bound value of $C_{max}$ can be approximated as the Li concentration of $Li_2O$, which is $8.11 \times 10^{22}$ cm$^{-3}$. Assuming a coating thickness of 1nm, and a calculated Li diffusivity in am-$Al_2O_3$ of $D = 5.94 \times 10^{-17}$ cm$^2$/s [37], then the largest current density one can obtain in steady state is only 0.0077 mA/cm$^2$ (where area is active electrode surface), which is only about 0.17 C rate (see Appendix III for the active area current density $J_{active}$ estimation corresponding to 1 C rate). If we further extend this analysis to some of the fastest rates explored on thicker coatings, say 10C on coatings of 100nm [47], we see that the possible steady state flux is about 6,000 times too small to be consistent with what is obtained in experiments. This analysis suggests that, given the low diffusivities of these materials, the electron-conducting model for a conformal coating, where Li transport can only be driven by its own concentration gradient, cannot provide a sufficient Li flux. We note that this analysis assumes that the calculated $D$ value used in our analysis is appropriate for the materials in the battery, which is uncertain (see Section IIIC). Thus the electron-conducing model cannot be totally ruled out by this analysis. However, given the poor agreement with experiments of our best present estimates, we assume this electron-conducting model is unlikely to be relevant for Li transport in the coatings considered in this work.

The second model assumes that the coating is an electronic insulator and a $Li^+$ space charge region exists inside the coating without any negative compensating charge. Based on the previous analysis of the electron-conducting model, the electric field must be the main driving force for Li transport (rather than concentration gradients) to obtain adequate current. In a steady state condition the current density $J$ is a constant with respect to the spatial coordinate $x$. Based on the general solution for current-voltage relationships in the space-charge limited regime,[48] we can calculate the potential drop across the coating, $\Delta V$, as $\Delta V(J,L) = \frac{2}{3}\sqrt{\frac{2k_BT}{q\varepsilon D}}J^{1/2}L^{3/2}$, where $T$ is temperature, $\varepsilon$ is the dielectric constant of the coating, and $L$ is the coating thickness. Assuming the coating is playing a significant role in the battery overpotential, which certainly seems to be the case for some of the thicker coatings [47] [49], then $\Delta V$ will be a significant portion of the observed battery overpotential. However, $\Delta V$ is proportional to $J^{1/2}$ and $L^{3/2}$. This trend doesn't match what has been found from previous experimental work in Refs. [47] [49] [50] [51] where the overpotential has an approximately linear relationship as a function of both $J$ and $L$. Another issue with this model is related to the magnitude of the electric field generated in the coating as a result of the $Li^+$ space charge. If a 1C rate current density (~0.046 mA/cm$^2$, see Appendix III) is flowing through the coating at room temperature, and we substitute the calculated am-$Al_2O_3$ diffusivity $D=5.94 \times 10^{-17}$cm$^2$/s [37], the electric field $E(x)$ exceeds the breakdown field of crystalline α-$Al_2O_3$ of 1.5 V/nm [52] (here we use the crystalline $Al_2O_3$ breakdown field to approximate that of the am-$Al_2O_3$) when the thickness is only $x=1$Å. If $x = 1$nm, the electric field will be ~4x higher than the breakdown field. Considering the above two factors, this space charge model is unlikely to be relevant for Li transport in nominally insulating ALD conformal coatings.

The final model we consider is the electrolyte model. This model is qualitatively consistent with previous experimental work in Refs. [47] [49] [50] [51] which find that overpotential is proportional to $J$ and $L$. Given the consistency of the electrolyte model with our present understanding of the origins and performance of the coatings, we will use this model for the analysis of the influence of Li diffusivity on current – voltage relationships in the rest of this work We discuss the electrolyte model, including the possible origins and nature of the compensating negative charges, in Section IID.

D.      Electrolyte model for coatings

We model the overpotential across a coating film as a function of the solubility and diffusivity of lithium ions. These calculations allow us to quantify the connection between the ability of the film to transport lithium ions and its performance.[22,53-55] This model relates film thickness, Li solubility, and Li diffusivity with overpotential at a given current, providing a useful qualitative guideline for what coating properties are necessary to maintain acceptably low losses in the battery. As discussed in Section IIC, we model the ionic conductivity as if the coating were an Ohmic electrolyte with Li all in the form of $Li^+$ and a compensating background negative charge that is immobile. Within this model the $Li^+$ concentration and electric field are constant within the coating and $Li^+$ diffusion is driven by the field. Within the electrolyte model the ionic conductivity due to Li diffusion is given as

$$\sigma = Cq\mu = (q^2/k_BT)DC \quad (2)$$

The quantity $C$ is the Li$^+$ concentration in the coating. We have used the Einstein relation, $D = (k_BT/q)\mu$, to relate the ionic mobility $\mu$ and the ionic diffusivity D of an ion of charge $q$ (for Li$^+$ ion $q = |e| = 1.602 \times 10^{-19}$ C). From Eq. (2) the overpotential, $\Delta V$, across a coating film of thickness $L$ when an electric current density $J$ passes through it can be calculated as[22,53-55]

$$\Delta V = JL/(q\mu C) = JLk_BT/(DCq^2) \quad (3)$$

Eq. (3) shows that $\Delta V$ is inversely proportional to Li$^+$ ion concentration $C$ and diffusivity $D$. The resistivity of the coating can be obtained as:

$$\rho = \Delta V/(JL) = k_BT/(DCq^2) \quad (4)$$

For our calculations of the resistivity the coatings we need values of $D$ and $C$ in Eq. (4). The $D$ values will be obtained from calculations in this work and in the literature. Within the electrolyte model $C$, the Li$^+$ concentration in the coating is controlled by the concentration of negative compensating charges in the coating. In order to estimate this concentration we again consider ALD Al$_2$O$_3$ films as a widely studied representative example. An estimation of the Li$^+$ concentration in the ALD am-Al$_2$O$_3$ coating can be obtained as follows. In the typical growth process of ALD Al$_2$O$_3$ thin film, H$_2$O is usually used as the oxygen precursor.[56] After the growth of the ALD Al$_2$O$_3$ film, atomic hydrogen is usually detected in the coating. Hydrogen stays in the coating in the form of H$^+$, i.e. protons. To balance the charge state of these protons and make the system charge neutral, there must exist donated electrons from the H atoms or some other defect states that can compensate these electrons. These compensating charges are the negative background charge indicated in the Ohmic electrolyte model. Based on Fig. 11 in Ref [56], it can be seen that the H atom percent varies from 6% to 22% without significant change of the O/Al ratio, which is always approximately equal to 1.5. These results indicate that there isn't a large number of Al vacancies compensating the H$^+$, which suggests electrons donated from H atoms are contained in the material. During the charging/discharging process, we assume that Li$^+$ will ion-exchange with H$^+$ (which leaves the coating and enters the electrolyte) and yield a Li$^+$ concentration equal to that of the original H$^+$. For the system to behave as an electrolyte rather than a conductor, it is necessary that the compensating negative electrons are immobile. We therefore assume these electrons are trapped in localized states created during the ALD process. Within this picture, the concentration of H$^+$ after the growth of the ALD coating qualitatively determines the maximum Li$^+$ concentration inside the coating during the following charging/discharging cycles. We take the H$^+$ concentration to be 14% based on the average value of the range 6% - 22% given in Ref. [56]. If we assume all of the H$^+$ is replaced by Li$^+$ in the charging/discharging process, the chemical formula of the system can be written as Li$_{0.81}$Al$_2$O$_3$, and the corresponding Li$^+$ concentration in the coating is about $1.52 \times 10^{22}$ atoms/cm$^3$. This concentration (in atoms per unit volume) can be obtained from the density of am-Al$_2$O$_3$, which we take as 0.0939 atoms/Å$^3$ or 53.248 Å$^3$ per Al$_2$O$_3$ formula unit [37] (here we assume no volume expansion after Li$^+$ exchanges with H$^+$ because the atom percent of Li$^+$ is small (~10%)). This Li+ concentration values will be used for $C$ with Eq. (4) to calculate the resistivity of the am-Al$_2$O$_3$ coating. We will also use this value for more general estimates for the crystalline and am-AlF$_3$ coatings discussed in this paper. Although approximate, this concentration is likely to provide a reasonable estimate for typical H$^+$ (and corresponding Li$^+$) concentrations in ALD grown films.

### III. Results and Discussions

#### A. Lithium interstitial defect migration in crystalline coatings

As discussed in Sections IIC and IID, within the electrolyte model the Li$^+$ ion is the relevant species of lithium existing in the coating materials, as this is the stable state of lithium expected when replaces the H$^+$ in the coating. Although the Li$^+$ state is forced by our presently adopted electrolyte model, it is of potential interest within other models to understand the Li energetics and solubility with respect to an external reference state for these materials. Therefore, an analysis of the Li energetics and solubility in terms of the host electronic structure in given in Appendix I. Within our present electrolyte model only Li$^+$ is present so we will consider the diffusion of the Li$^+$ ion, focusing on migration via a nearest-neighbor interstitial hopping mechanism. All migration pathways of the lithium ion with open space and relatively short hopping distance are searched using the CI-NEB method and the corresponding values of the migration energy barrier, $E_m$, have been calculated. A complete discussion of the different Li diffusion pathways and comparison of Li migration barriers for the different coating materials is contained in Appendix II. The migration barriers for minimum-energy pathways are given in Table 1 with the corresponding estimated values of the diffusivity and mobility at 300 K. Similar CI-NEB calculations (not shown) were also performed for the case of Li$^0$ diffusion as a check in all the crystalline coatings. However, there are no significant differences in the calculated migration barrier when Li$^0$ versus Li$^+$ is used as the diffusing species. This result is consistent with the fact the Li$^0$ will ionize to Li$^+$ in the coating with its electron delocalized from the Li, resulting in nearly identical behavior of the diffusion of Li$^0$ and Li$^+$ in these materials. It is likely that a similar situation occurs in amorphous coatings. Therefore, in order to calculate the diffusivity and mobility for the case of Li$^+$ ion diffusing in am-Al$_2$O$_3$ and am-AlF$_3$, we simply reuse the distribution of migration barriers for the Li$^0$ case already reported in Ref. [37]. We revisit our discussion of the amorphous coatings in Section IIIC.

In recently published simulation work from Kim, et al. [57] similar calculations of the interstitial Li diffusion in crystalline Al$_2$O$_3$ and SiO$_2$ were performed. Encouragingly, the migration pathways obtained by Kim, et al. and in this work are quite similar. However, Kim, et al.'s calculated migration barriers were $E_m$ = 0.162 eV for SiO$_2$ and $E_m$ = 1.020 eV for Al$_2$O$_3$, which are significantly lower than our barrier values of 0.276 eV and 2.498 eV for SiO$_2$ and Al$_2$O$_3$, respectively. We believe that the discrepancy is largely due to the use of full relaxation of all images during the NEB calculations by Kim, et. al.. This full relaxation differs from the approach used for calculations in this work, which kept the volume and cell parameters fixed during the CI-NEB calculation, although the cell-internal coordinates were relaxed. If we fully relax the cell parameters in a manner analogous to Kim, et al. we obtain $E_m$ = 1.146 eV for Al$_2$O$_3$, much closer to their calculated value. We believe that constraining the volume and cell parameters during the relaxation, as done in our study, is more accurate as it avoids strong coupling of the cell size and shape to the migrating atom and its images in the periodic supercells.

**B. Lithium concentration and migration energies in amorphous Al$_2$O$_3$ and AlF$_3$**

The coatings put down with Atomic Layer Deposition (ALD) are likely to be in an amorphous structure, as many reports in the literature show coatings that were in an amorphous form.[2,4,17,18] Moreover, as mentioned previously, some of the authors of this work have recently found from their simulations that amorphous forms of coatings such as am-Al$_2$O$_3$ and am-AlF$_3$ have migration barriers lower than their crystalline counterparts.[37] For these amorphous materials, Li migration barriers are not just specific single values for certain insertion sites or diffusion pathways in the crystals as presented in Sections IIIA; rather, they are a distribution of values over certain ranges (see Ref. [37] for more detail). Here we describe how we model the Li concentration and diffusivity in am-Al$_2$O$_3$ and am-AlF$_3$ to allow application of Eq. (4).

The estimation of the lithium concentration in the amorphous coatings is done following the approach in Section IID, which yielded a value of 1.52x10$^{22}$ Li/cm$^3$. The estimation of Li atom diffusivities in the two amorphous coatings is done by fitting Eq. (1) to the results of kinetic Monte Carlo simulations of Li hopping in the amorphous structure given in Ref. [37]. This fitting yields values of $D_0$ = 1.09x10$^{-4}$ cm$^2$/s (am-Al$_2$O$_3$) / 7.69x10$^{-5}$cm$^2$/s (am-AlF$_3$), $E_m$ = 0.73 eV (am-Al$_2$O$_3$), / 0.65 eV (am-AlF$_3$), and $D$ = 5.94x10$^{-17}$ cm$^2$/s (am-Al$_2$O$_3$), / 9.26x10$^{-16}$ cm$^2$/s (am-AlF$_3$) at T=300K. These values are included in Table 1 with the corresponding mobilities estimated using the Einstein relation for comparison to those of Li$^+$ diffusion in crystalline coatings.

In recent work Jung, et al.[38] reported that the intercalated Li may react with Al$_2$O$_3$ first and form am-Li$_{3.5}$Al$_2$O$_3$. The solubility of Li in this new phase is about 3.03x10$^{22}$/cm$^3$ (this number can be calculated from the supporting information of Ref. [38] where they give the volume expansion due to Li insertion as $V/V_0$ =2.1, where $V$ is the volume of Li$_{3.5}$Al$_2$O$_3$ and $V_0$ is the volume of pure Al$_2$O$_3$, for which we use the values given in Section IIC). This is, for our purposes, quite close to the Li solubility we estimate for the am-Al$_2$O$_3$ (1.52x10$^{22}$/cm$^3$). Therefore, this new am-Li$_{3.5}$Al$_2$O$_3$ material doesn't greatly enhance the solubility compared to our calculated value for the am-Al$_2$O$_3$. However, the calculated diffusivity of Li in am-Li$_{3.5}$Al$_2$O$_3$ (~ 7.1x10$^{-10}$ cm$^2$/s) is predicted to be much higher (by approximately seven orders of magnitude) than the diffusivity of Li in am-$Al_2O_3$ (~$5.9 \times 10^{-17}$ cm$^2$/s). This difference comes from two parts: (1) $D_0$ in am-$Li_{3.5}Al_2O_3$ is $1.5 \times 10^{-3}$ cm$^2$/s which is ten times larger than $D_0$ in am-$Al_2O_3$, and (2) the migration barrier in am-$Li_{3.5}Al_2O_3$ is 0.35 eV lower compared with the barrier in am-$Al_2O_3$. Jung, et al. also showed the diffusivity of Li in the relatively dilute Li case of $Li_{0.2}Al_2O_3$, which was predicted to be a $D$ value of $1.1 \times 10^{-14}$ cm$^2$/s. We can compare this diffusivity value with the diffusivity of $5.94 \times 10^{-17}$ cm$^2$/s we estimated from Ref. [37], where the Li content is $Li_{0.00625}Al_2O_3$. Both of these Li concentrations might be reasonably considered dilute and therefore the values are expected to be similar. The values differ by about a factor of 200x, which is reasonable considering the concentration dependence of the diffusivity and the possible DFT errors. Another recent published work[20] also shows that the amorphous $LiAlO_2$ thin film (another composition in the Li-Al-O ternary with high Li content) has a much higher Li diffusivity compared with am-$Al_2O_3$. The calculated diffusivity matches quite well with experimental measurements yielding a Li diffusivity of approximately $10^{-11}$ cm$^2$/s. This result further indicates the possibility that alloying of the am-$Al_2O_3$ with Li to form an amorphous Li-metal oxide compound may provide a fast Li conducting pathway.

### C. Discussion

The goal of the current work is to understand lithium diffusivity in crystalline and amorphous coatings and their impact on the electrochemical performance, specifically the overpotential caused by coating films. Let us consider how the lithium diffusivity of the coatings compares to those of typical solid-state materials for Li-ion batteries. A list of common materials related to Li-ion batteries and their Li diffusivities and migration barriers are given in Fig. 2. Here, for the sake of comparison, Eq. (1) is used to estimate these diffusivity values from the corresponding values of DFT-calculated migration barriers collected from literature and with the same $D_0 = 2.5 \times 10^{-2}$ cm$^2$/s as used for the crystalline insulator coatings studied in this work (see Table 1). It is obvious from our calculations that the Li$^+$ ion diffusivities at room temperature of the crystalline coatings in question (except for α-quartz $SiO_2$) and even the am-$Al_2O_3$ and am-$AlF_3$ coatings, are many orders of magnitude lower than that for typical electrode materials,[22] such as olivine-structured $LiFePO_4$, layer-structured $LiCoO_2$, spinel-structure $LiMn_2O_4$ and graphitic carbon anode materials.[58] The low values of Li$^+$ ion diffusivities for these insulator coatings is largely due to the relatively high range of Li$^+$ ion migration energy barriers, where α-quartz $SiO_2$ is an exception with a Li$^+$ ion migration energy barrier comparable to those for cathode materials, at least along the <001> direction.[59] The crystalline binary oxide coatings considered in this work (other than α-quartz $SiO_2$) also facilitate much slower lithium diffusion than some other binary oxide coatings, such as ZnO[60] and $TiO_2$[61], and slower Li transport than solid electrolyte coatings such as perfect and imperfect (i.e., O defected and N or Si substituted) $Li_3PO_4$ crystals.[62-64] It is also worth noting that the lithium diffusivities of the crystalline insulator coatings other than α-quartz $SiO_2$ are much lower than those for $Li_2CO_3$ and $Li_2O$, and somewhat lower than that for LiF, as these are three main solid-state components of the inner dense layer of the solid-electrolyte interphase (SEI) formed on carbonaceous anode surfaces.[65] The am-$Al_2O_3$ and am-$AlF_3$ coatings are also generally slower diffusers that these SEI phases, although they are comparable to LiF. A fundamental difference between the perfect insulator coatings in this work and the lithium-transport components of the SEI inner dense layer and solid electrolytes in Li-ion batteries compared in this section is that the latter are lithium compounds while the former are not. Consequently, the lithium transport mechanisms and ionic defect carriers may be quite different: specifically, transport is probably only by lithium interstitials in the coating materials but both lithium sublattice vacancy, interstitialcy, and interstitial mechanisms in the Li compounds may contribute to their ionic transport properties. The above observations suggest that all the crystalline phases other than α-quartz $SiO_2$ are likely to be too poor at Li transport to be practical coating materials, regardless of any additional issues associated with dissolving enough Li to allow a significant Li flux. However, it is difficult to judge what Li diffusivities and solubilities are actually needed to enable adequate transport for a nanoscale coating without a more detailed model of how the small diffusion distances couple to current and overpotential in the battery. Here we present results on overpotentials and resistivities predicted from our Ohmic electrolyte model using Eqs. (3) and (4).

Table 1 gives the resistivities predicted by our Ohmic electrolyte model from the estimated Li solubility and calculated diffusion coefficients for each material studied. The Li$^+$ concentration in am-$AlF_3$ is approximated to be equal to that in am-$Al_2O_3$, which was estimated in Section IID. To help understand the coupling of Li solubility and diffusivity to overpotential more intuitively, Fig. 3 presents the plot of the overpotential across the coating ($\Delta V$) vs. $C$ and $D$ for a general coating of thickness 1 nm on a cathode with a current density of $J_{active} = 0.046$ mA/cm$^2$ at room temperature (300 K). The current density given here corresponds to a cycling rate of 1 C for a real Li-ion battery with a $LiCoO_2$ cathode. Refer to Appendix III for the discussion about how this current density was obtained. This current density is through the coating layer and given per unit area of coating over the Li-intercalation active cathode surface. This current density will be denoted as $J_{active}$, as it is normalized by the cathode surface area active for Li transport, and it is to be distinguished from the more common $J_{geom}$, which is the current density normalized per unit geometric area of the cathode disk, which is based on the area of the cell normal to the Li transport direction. The overpotential data in Fig. 3 is calculated from Eq. (3). We note from Eq. (3) that the data in Fig. 3 can be shifted to arbitrary current and thickness by simple linear scaling of the voltage with those values relative to the values used here. In the following we focus on what is required to maintain an overpotential of < 0.1 V across the coating, as this is a reasonable upper limit for what might be tolerable in a battery. Within the validity of the model represented in Fig. 3 we see that to maintain an overpotential of $\Delta V < 0.1$ V at ~ 1 C and $T = 300K$ through a conformal coating, even with perfect $Li^+$ solubility of $C \approx 10^{23}$ $cm^{-3}$, one would need a diffusivity larger than $\approx 10^{-14}$ $cm^2/s$ and $\approx 10^{-13}$ $cm^2/s$ for a 1 nm and 10 nm film, respectively, which corresponds to a migration barrier less than about 0.74 eV and 0.68 eV, respectively, using $D_0 = 2.5 \times 10^{-2}$ $cm^2/s$. We also see that even if the diffusivity of the coating was as fast as that of a high-performing cathode material such as $LiCoO_2$, ($D \sim 10^{-7}$ $cm^2/s$) its solubility would need to be of the order of $10^{16}$ $Li/cm^{-3}$ in order to achieve $\Delta V < 0.1$ V. The constraints suggested by the above model immediately imply that all the crystalline materials except α-quartz $SiO_2$ have barriers that are too high to allow reasonable performance, even with just a 1 nm coating. This result relies on the significant assumptions that lead to Eq. (3), but are consistent with the observations that these materials are poor diffusers compared to other materials that successfully transport Li in a battery (as shown in Fig. 2). However, α-quartz $SiO_2$ is an interesting exception. For the slower direction it still provides slow but possibly practical diffusion (for very high Li solubilities), and along <001> it provides very rapid diffusion. The fast diffusion along <001> is consistent with previous experiments on alkali atoms in $SiO_2$ (see Appendix II). Thus α-quartz $SiO_2$, although it might need to be oriented to allow transport along the <001> direction, could potentially provide a very fast transport conformal crystalline coating material.

A number of previous experimental studies have been performed on $SiO_2$ coating layers on different types of cathodes. $SiO_2$ has been coated on the layered structure materials $LiNi_{0.8}Co_{0.15}Al_{0.05}O_2$[33] and $LiNiO_2$,[66] olivine $LiFePO_4$,[35] monoclinic $Li_3V_2(PO_4)_3$[34] and spinel $LiNi_{0.5}Mn_{1.5}O_4$.[67] All of these examples reported enhanced structural stability, improved capacity retention and better electrochemical performance of the cathode material associated with the $SiO_2$ coating. The last two works[34,67] also proposed that the $SiO_2$ might be a good HF scavenger[15] through the reactions $SiO_2 + 4HF \rightarrow SiF_4 + 2H_2O$ and $SiO_2 + 6HF \rightarrow H_2SiF_6 + 2H_2O$.[68] This effect may be another reason to explain why the $SiO_2$ coating can protect cathodes and improve the battery performance.

The amorphous materials studied here are, in general, significantly better Li transporters than their crystalline counterparts. They allow for a high Li concentration of $1.52 \times 10^{22}$ /$cm^3$ (see Section IID) and relatively low migration barriers. Given the above estimated diffusivities and the concentration of $Li^+$ ions in the am-$Al_2O_3$ and am-$AlF_3$ coatings we can substitute them into Eq. (3) and calculate the overpotential across a 1 nm thick conformal amorphous coating at an approximately 1C charging rate. We find that the overpotentials of am-$AlF_3$ and am-$Al_2O_3$ are 0.051 V and 0.82 V respectively. To quantify how close this performance is to what might be needed, we compare these calculated results directly to resistance properties estimated from nominally conformal coatings. Note that because we wish to focus on at least nominally conformal coatings, this limits us to coatings deposited by Atomic Layer Deposition. While both $AlF_3$ and $Al_2O_3$ have been widely studied, $Al_2O_3$ is the only one of these two materials which, to our knowledge, has been coated using ALD on cathodes. In Table 1 we see that the effective resistivity of the am-$AlF_3$ and am-$Al_2O_3$ coatings can be calculated from the models in this paper as 114 MΩm and 1789 MΩm (where MΩm = $10^6$Ωm), respectively. Some experimental estimates for relevant nominal $Al_2O_3$ coating resistivity can be obtained from Refs. [47,49-51], which studied ALD deposited $Al_2O_3$ coatings on $LiCoO_2$ and NMC cathodes. The details of the analysis used to find the coating resistivity values are summarized in Appendix III. The range of the estimated am-$Al_2O_3$ resistivities fitted from previous experiments is 7.8 MΩm to 913 MΩm. While the structure of the ALD $Al_2O_3$ coating in the operating battery is not totally clear, it is expected to be somewhat amorphous and react to at least partially fluorinate,[69] so comparison to am-$Al_2O_3$ and am-$AlF_3$ is reasonable. Comparison to our predictions show that our am-$Al_2O_3$ calculation (calculated am-$Al_2O_3$ resistivity is 1789 MΩm) is higher than the range of our fitted resistivities (7.8 to 913 MΩm) based on previous experiments and our am-$AlF_3$ calculation (calculated am-$AlF_3$ resistivity is 114 MΩm) falls in the range. Our predicted am-$Al_2O_3$ resistivity is about 2× higher than the maximum (913 MΩm) and about 230× larger than the minimum (7.8 MΩm) of the experimental range, suggesting the model is more consistent with the maximum fitted values. Given the uncertainties in the modeling and the extraction of experimental data (these uncertainties are discussed further below), the discrepancy between the largest experimental resistivity (913 MΩm) and the modeling resistivities for either am-$Al_2O_3$ or am-$AlF_3$ are almost certainly within their combined uncertainty. In general the range of fitted experimental resistivities is somewhat too large to provide a highly quantitative restriction. Therefore, for completeness and to guide future work, it is important to consider possible sources of quantitative disagreement between our model and experimental results, which we discuss in the following paragraphs.

Here we consider possible sources of errors in the $D$ values we estimated from the experiments. We note that it is possible that in the experiments other Li transport paths besides the direct bulk transport of Li through the coating could be available, e.g. pinholes or uncoated regions of the cathode, which would lead to incorrect and low resistivity estimates for the coating. In addition, the experimental analysis is quite approximate, and could easily yield factors of two or perhaps more from use of approximate linear fits to approximate overpotentials and errors in the estimated effective active surface area. More broadly, our connection between the experimental resistivity and the Li diffusivity is through an Ohmic model that is appropriate for an electrolyte system (see Sec. IID), and it is possible that this model does not rigorously apply for these coatings. However, this model is consistent with the linear potential and coating thickness relationship seen in many experiments (see Appendix III). If we assume that our estimates of resistivity and Li diffusion coefficients extracted from experiment are reliable, then major discrepancies are likely due to either errors in the model approach or differences between the material being modeled and the real experimental system. We now consider each of these in turn.

One possible source of error in the model resistivity calculation may come from the calculation of the Li migration barrier. We note that the Li migration barriers for the amorphous materials had to be extracted from a complex multiscale *ab initio* and kinetic Monte Carlo simulation in Ref. [37], which could lead to errors. These barriers would have to be overestimated by about 141 meV (18.5% of the calculated am-$Al_2O_3$ migration energy) to yield results consistent with the lowest value from the experiments and by 18 meV (2.4% of the calculated am-$Al_2O_3$ migration energy) to yield results consistent with the highest value from the experiments. This 18.5% error is significantly larger than the errors seen in models done using similar techniques for $LiAlO_2$[20] (where the discrepancy with experimental diffusivity corresponded to only about 20 meV in an Arrhenius expression, consistent with less than a 4% error in barrier assuming the error is all due to the barrier). The 2.4% error is not unreasonable for a DFT migration barrier calculation, suggesting we are within DFT energy errors compared with the highest values of the resistivity. It is further possible, and even likely, that the dilute Li migration energy values used in this work would be altered at the significant Li concentrations that may be present in the amorphous coating.

Another possible source of error is that the estimation of $Li^+$ concentration derived from our coating electrolyte model in Section IID may have errors. The range of $H^+$ concentrations observed suggests that a factor of two error in our estimated concentration could easily occur, and different synthesis methods might lead to larger differences. Furthermore, the model proposed in Section IID for what controls the Li concentration in quite speculative, and further study is needed to assure its validity.

We now consider the question of possible discrepancies due to the material being modeled as pure bulk am-$Al_2O_3$ being different from the actual material in the operating battery. Specifically, in the battery the coating material may be a mixed fluoride and oxide due to reaction with fluorine.[7,15] The calculated am-$AlF_3$ coating yields a predicted resistivity (114 MΩm), lower than the predicted resistivity of am-$Al_2O_3$ (1789 MΩm) due to the higher Li diffusivity of am-$AlF_3$. Given that am-$AlF_3$ is a faster diffuser than am-$Al_2O_3$, it is possible that fluorination may increase Li diffusivity compared to pure am-$Al_2O_3$, lowering the resistance of the material. Such a process could help explain the somewhat lower values of resistivities extracted from the experiments compared to the theoretical predictions for pure $Al_2O_3$. Other differences between the model and experimental material may be that the measured material is altered by alloying with Li,[38] may have an amorphous structure more open than produced by the rapid liquid quench technique used in the modeling in Ref. [37], may interact significantly with the cathode (particularly likely for very thin coatings),[70] or may be highly defected in ways that alter Li transport. In particular, based on the results of Jung, et al,[38] Li may react with am-$Al_2O_3$ and form am-$Li_{3.5}Al_2O_3$, which has a very high predicted Li diffusivity (see Table 1).[38] From Eqs. (3) and (4) and the predicted diffusivity from Jung, et al,[38] we estimate the resistivity for $Li_{3.5}Al_2O_3$ as $7.4 \times 10^{-5}$ MΩm and the overpotential across a 1 nm thick conformal amorphous coating at 1C charging rate to be just $\Delta V = 0.34 \times 10^{-7}$ V. It should be noted that within the PBE-GGA DFT approach of Ref. [38] the am-$Li_{3.5}Al_2O_3$ phase is predicted to have zero band gap (see p.10 of supplemental information of Ref. [38]). If this phase is in fact metallic then the Ohmic electrolyte model used here is not applicable and the impact of the material on overpotential must be modeled following the electron-conducting model in Section IID, which is beyond the scope of the present work.

However, if we assume that this phase works by the electrolyte model described in Section IID and has the predicted diffusivity from Ref. [38] then it is actually too fast of a Li diffuser to explain the significant resistivity observed experimentally in Refs. [47,49-51]. However, some other Li-Al-O compound may form and provide a more intermediate Li diffusivity consistent with observations, or perhaps a fast diffusing lithiated phase forms only over very small regions. Analogous arguments can be made concerning the fast diffusing amorphous $LiAlO_2$ films studied by Park et. al.[20] Finally, we note that the Li transport behavior of the coating material has been modeled as homogenous and identical to an approximately infinite material. For thin coatings, fluctuations in the local amorphous structure may lead to significant variation in effective Li concentration and/or diffusion coefficient, and thereby enhance or retard Li transport through some regions. Such fluctuations are beyond the scope of the present study but are an area of potential interest for future work.

Overall, these results imply that the crystalline phases $\alpha$-$AlF_3$, $\alpha$-$Al_2O_3$, m-$ZrO_2$, c-MgO generally cannot be practically used as conformal crystalline coatings at even 1 nm thick, but that $\alpha$-quartz $SiO_2$ might be a practical crystalline material. Furthermore, our calculated resistivity of pure am-$Al_2O_3$ is higher than the range of our fitted resistivities from previous experiments (2x higher than the maximum, 230x higher than the minimum) and pure am-$AlF_3$ falls within the range. However, the large uncertainties of the fitted experimental results make it difficult to assess the quantitative agreement between experiments and our present model, and many possible sources of quantitative errors exist. Nonetheless, the results do suggest that some of the very fast diffusing Li-containing $Al_2O_3$ phases that have been proposed are not consistent with the high resistivities observed unless they allow only very small area pathways from the electrolyte to the cathode or assume a different model for the coating transport than our electrolyte model.

It is worth noting that non-conformal coatings also appear to be successful, for example, the small-particle-on-large-particle or the rough coatings reported in Refs. [16,36]. The coating in these non-conformal cases may play the role to reduce the direct contact area to some extent, but not entirely, between electrolyte and cathode. Such non-conformal coatings may also preferentially bind to reactive sites and suppress electrolyte oxidation or cathode corrosion, although their mechanisms of enhancing performance are not well established (as discussed in Section I).

Based on our $Li^+$ solubility discussion (Section IID and Appendix I), proper defect control may help to improve $Li^+$ diffusivity and conductivity, as reported for imperfect $Li_3PO_4$ solid electrolytes[64] and for $Li_2CO_3$ in the SEI layer.[71] More specifically, as we discussed in the estimation of $Li^+$ solubility in Section IID, if we can create a higher concentration of negative compensating charge in the system, then the $Li^+$ solubility will be higher and it will enhance the Li transport across the coating. However, it should be noted that defects could facilitate electron transport in an otherwise insulating coating, which could enable electrons to leak through the coating and potentially harm not only the stability of the coating itself but also that of both cathode and electrolyte in terms of redox reactions among their species.

## IV. Conclusions

We have carried out first-principles calculations based on Density Functional Theory to examine the diffusivity of Li in a number of idealized insulator cathode coatings in their room temperature and pressure stable crystalline structures ($\alpha$-$AlF_3$, $\alpha$-$Al_2O_3$, m-$ZrO_2$, c-MgO, $\alpha$-quartz $SiO_2$) and adapted previously published results[37,38] for selected amorphous structures (am-$Al_2O_3$, am-$Li_{3.5}Al_2O_3$, and am-$AlF_3$). We assume that the coating behaves like an electronically insulating but ionically conducting electrolyte for Li transport, and we use an Ohmic electrolyte model to estimate the coating resistivities. We find that $Li^+$ ions diffuse quite slowly in the crystalline coatings, with a migration barrier $E_m$ larger than 0.9 eV in crystalline $\alpha$-$AlF_3$, $\alpha$-$Al_2O_3$, m-$ZrO_2$, and c-MgO. We show by comparison to other Li transporting materials in batteries and a simple Ohmic electrolyte model that these materials cannot provide adequate Li transport to serve as practical conformal coatings. Among the crystalline materials studied, $\alpha$-quartz $SiO_2$ emerged as a particularly interesting material, with generally low Li formation energies and $Li^+$ migration barriers of just $E_m = 0.736$ eV along <100> and $E_m = 0.276$ eV along <001>. The low migration barrier for pathways along the <001> direction of $\alpha$-quartz suggests a diffusivity of $5.8 \times 10^{-7}$ cm$^2$/s at room temperature, making an oriented $\alpha$-quartz coating potentially a fast Li conductor. Combined with its high Li solubility compared to the other crystalline materials, $\alpha$-quartz $SiO_2$ emerges as interesting for further study. We further predict, based on previous calculations[37], that am-$Al_2O_3$ and am-$AlF_3$ are able to dissolve significant amounts of Li and are faster Li diffusers than their crystalline counterparts. Our Ohmic electrolyte model predicts that the calculated resistivity of pure am-$Al_2O_3$ is higher than the maximum value of the experimentally extracted resistivities and the calculated resistivity of pure am-AlF$_3$ falls within the experimental range. However, due to the large uncertainties it is difficult to achieve a highly quantitative assessment of our model compared to experiments. Furthermore, there are a number of possible sources of error between the modeling and experiments, including: the extraction of resistivity values from the experiments, incorrect assumptions or values in the electrolyte model, and differences between the materials in the model and those in the active battery.

This work develops an integrated approach to predicting coating overpotentials from atomistic simulations and fundamental coating properties such as the Li diffusion coefficient and solubility. This model is expected to be useful for future exploration of coatings and our successful prediction of overpotentials within the experimentally observed range helps validate the approach. The model suggests that electrode coatings of oriented α-quartz SiO$_2$ and am-AlF$_3$ are of interest as they have significantly faster diffusion than am-Al$_2$O$_3$. The comparisons of our model to experiments for ALD deposited am-Al$_2$O$_3$ suggest that, assuming they remain insulating, fast diffusing am-Li$_{3.5}$Al$_2$O$_3$ [38] and crystalline LiAlO$_2$ [20] phases do not form in the battery to enough of an extent to provide dominant Li transport pathways in the experiments to date.

## AUTHOR INFORMATION

**Corresponding Author**

*Dane Morgan. E-mail: **ddmorgan@wisc.edu**.**Author Contributions**

The manuscript was written through contributions of all authors. All authors have given approval to the final version of the manuscript.

## ACKNOWLEDGEMENTS

The authors gratefully acknowledge funding from the Dow Chemical Company and helpful conversations with Mark Dreibelbis, Brian Goodfellow, Thomas Kuech, and Robert Hamers. Computations in this work benefitted from the use of the Extreme Science and Engineering Discovery Environment (XSEDE), which is supported by National Science Foundation grant number OCI-1053575.11

**Table 1** Predicted migration barriers of the minimum-energy pathways and corresponding estimated diffusivities, mobilities (from Eq. (3)) at 300 K for crystalline coatings. The effective migration barriers and the associated diffusivities, mobilities, and resistivities at 300 K for am-$Al_2O_3$, am-$Li_{3.5}Al_2O_3$ and am-$AlF_3$ are also included for comparison. The $Li^+$ solubility in all systems is approximated to be equal to that in am-$Al_2O_3$ (see Section IIID).. The approximate resistivities of am-$Al_2O_3$ obtained from experiment are from 7.8 MΩm to 913 MΩm (see Appendix III).

| Material | Migration barrier (eV) | Diffusivity ($cm^2$/s) | Mobility ($cm^2$/(V-s)) | Resistivity (MΩm) |
|---|---|---|---|---|
| $Al_2O_3$ | 2.498 | $2.7 \times 10^{-44}$ | $1.06 \times 10^{-42}$ | $3.9 \times 10^{30}$ |
| $AlF_3$ | 0.929 | $6.2 \times 10^{-18}$ | $2.4 \times 10^{-16}$ | $1.7 \times 10^4$ |
| MgO | 1.419 | $3.6 \times 10^{-26}$ | $1.4 \times 10^{-24}$ | $2.9 \times 10^{12}$ |
| $ZrO_2$ | 0.962 | $1.7 \times 10^{-18}$ | $6.7 \times 10^{-17}$ | $6.2 \times 10^4$ |
| $SiO_2$ <001> | 0.276 | $5.8 \times 10^{-7}$ | $2.2 \times 10^{-5}$ | $1.82 \times 10^{-7}$ |
| $SiO_2$ <100> | 0.736 | $1.1 \times 10^{-14}$ | $4.1 \times 10^{-13}$ | 9.6 |
| am-$Al_2O_3$[37] | 0.73 | $5.9 \times 10^{-17}$ | $2.24 \times 10^{-15}$ | 1789 |
| am-$AlF_3$[37] | 0.65 | $9.3 \times 10^{-16}$ | $3.52 \times 10^{-14}$ | 114 |
| am-$Li_{3.5}Al_2O_3$*[38] | 0.38 | $7.1 \times 10^{-10}$ | $2.69 \times 10^{-8}$ | $7.4 \times 10^{-5}$ |

* This material has been predicted to be metallic by density functional theory calculations.[38] If it is metallic then the Ohmic electrolyte model (Eq. (3)) and associated resistivity determined here are not applicable.

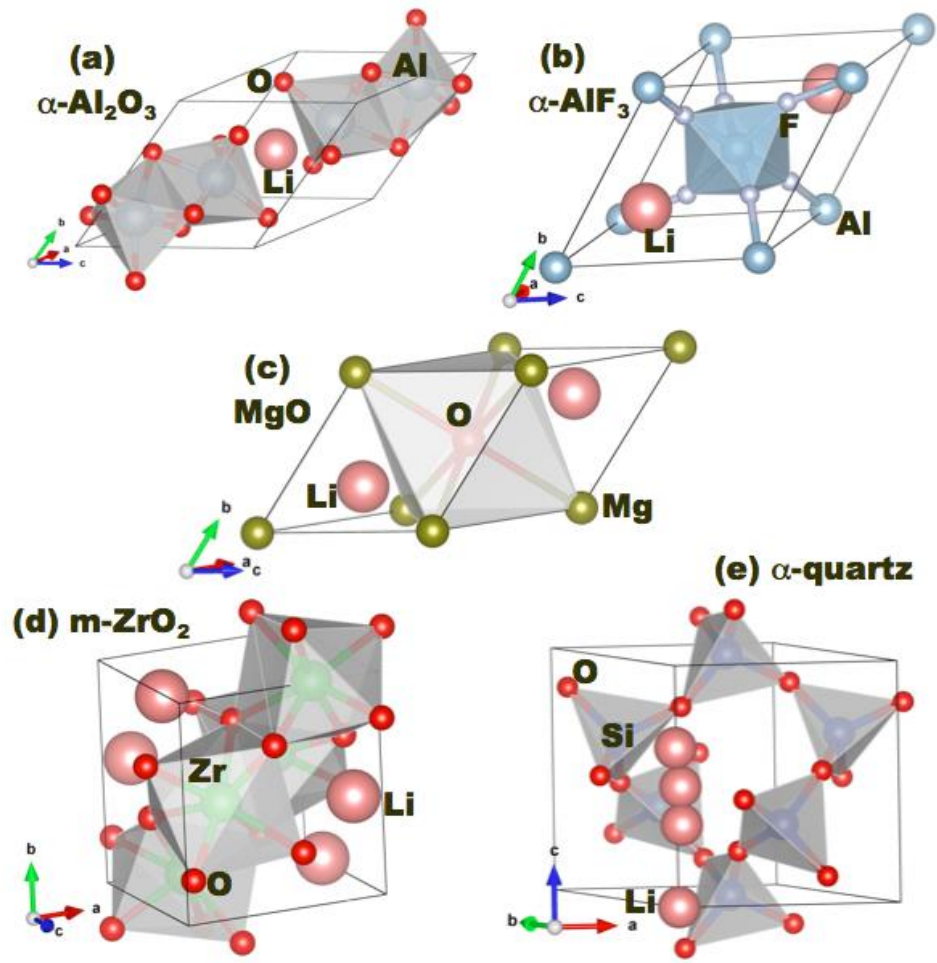

**Fig. 1** Primitive cells of the coating materials studied in this work. The positions of lithium interstitial sites in the cells are indicated.

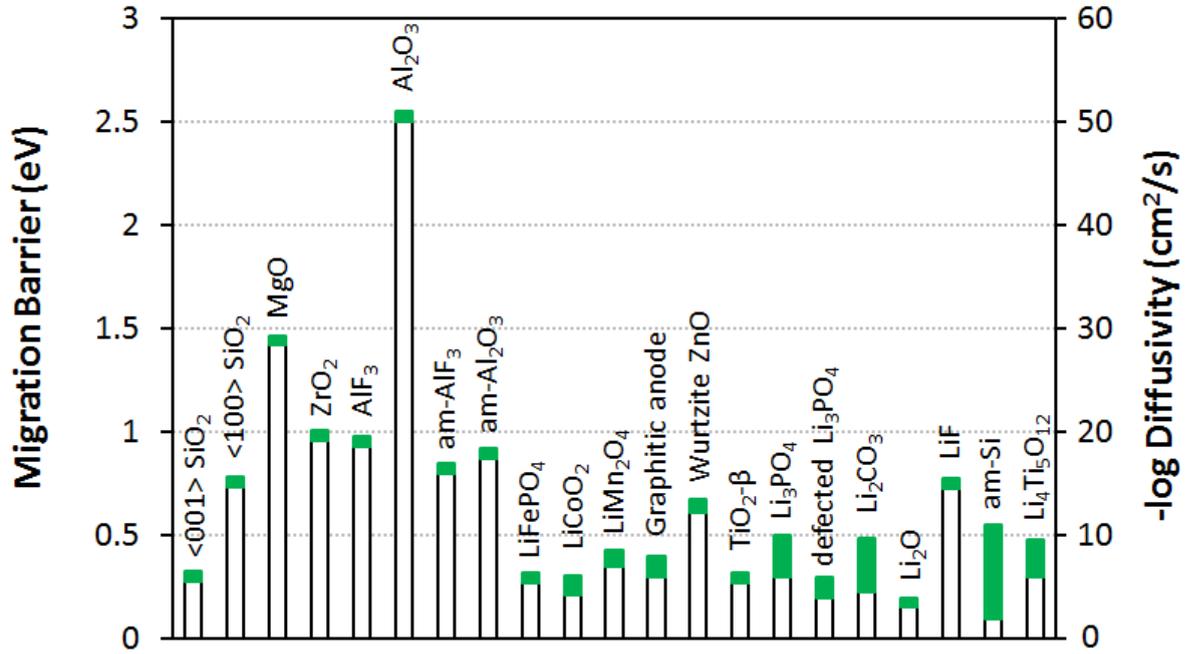

**Fig. 2.** Li migration barrier $E_m$ and diffusivity $D$ at 300K for numerous materials related to Li-ion batteries. The values of $E_m$ and $D$ for the coatings studied in this work are from Table 1 and for the other materials are determined with Eq. (1) using the $D_0$ values used for the crystalline phases in this work ($D_0 = 2.5\times10^{-2}$ cm$^2$/s) and the migration energy barriers collected from literature: Ref. [22] for LiFePO$_4$, LiCoO$_2$, and LiMn$_2$O$_4$; Ref. [37] for am-Al$_2$O$_3$ and am-AlF$_3$; Ref. [58] for graphitic anode; Ref. [60] for ZnO; Ref. [61] for TiO$_2$-β phase; Refs. [62-64] for Li$_3$PO$_4$ and defected Li$_3$PO$_4$; Ref. [63] for Li$_2$CO$_3$, Li$_2$O, and LiF; Ref. [72] for amorphous Si and Refs. [73,74] for Li$_4$Ti$_5$O$_{12}$. The green bars represent the approximate range of migration barrier and diffusivity values based on literature values and typical errors on DFT migration energy barriers (taken to be ±50 meV).

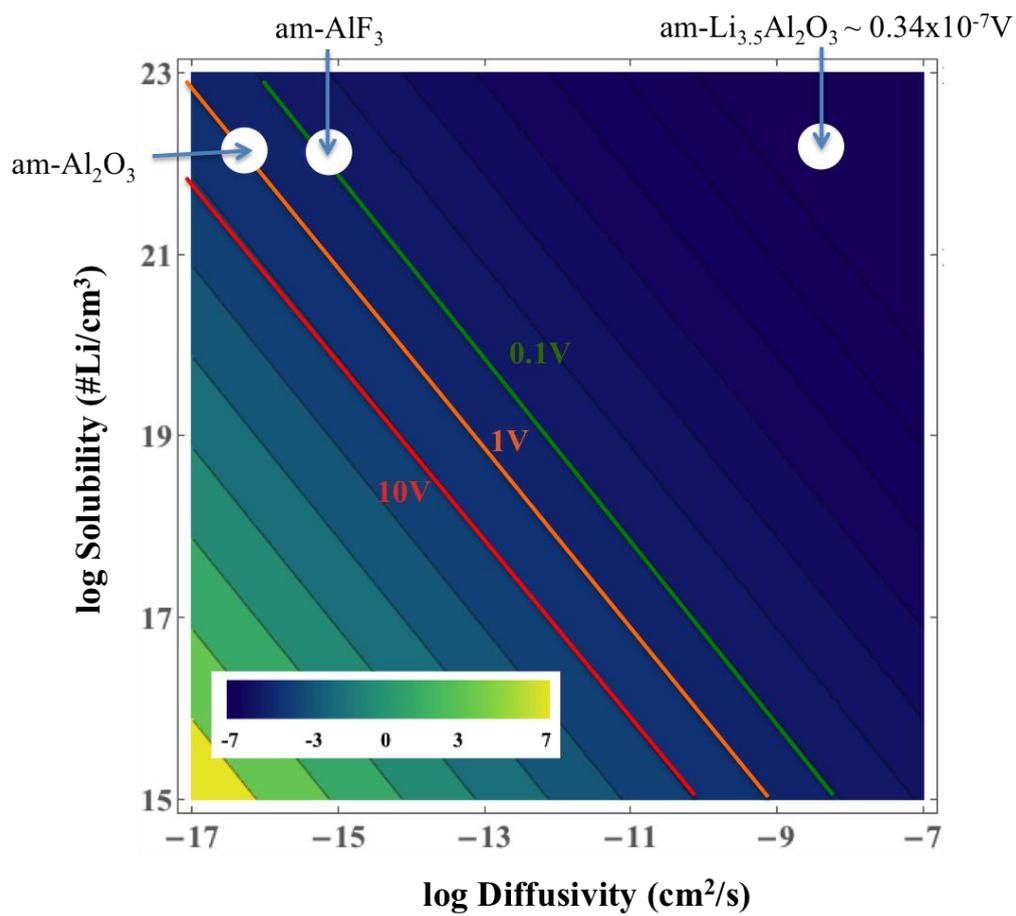

**Fig. 3.** Solubility and diffusivity dependence of potential drop across a room temperature thin coating film of $L = 1$ nm for a current of density $J_{active} = 0.046$ mA/cm$^2$ of active surface (approximately a 1C rate, as discussed in Appendix III). The values of the potential drop are calculated from Eq. (3).

**Appendix I: Lithium formation energetics in crystalline and amorphous coatings at equilibrium**

The chemical processes controlling the Li concentration proposed in Section IID are somewhat speculative, and other mechanisms may play a role. In particular, the Li concentration in the coating may be controlled by equilibrium with the anode, electrolyte, or cathode rather than a charge balance established during synthesis. In this appendix, we aim to provide further insight on the physics of Li interaction with these coating materials by using DFT calculations to explore the Li defect chemistry. The goal with these calculations is to better understand the behavior of Li insertion into these materials when both Li and its electron can insert into the coating and freely interact, including the solubility of Li under such conditions. In these equilibrium defect calculations, the solubility is not set by the available compensating negative charge in the coating material (as in the steady-state electrolyte coating model described in Section IIC and IID), but rather is dictated by the equilibrium of Li defect formation relative to a Li source. The source of Li is some external chemical reservoir, for which we choose Li metal. One may then simply shift all Li insertion energies obtained here to any arbitrary reference state, such as a particular cathode material of interest, by using the relative energies of the new reference state to Li metal.

**Computational methods for lithium defect formation**

All systems considered in this study are insulating in their undefected form, therefore we use the hybrid exchange and correlation functional of Heyd, Scuseria and Ernzerhof (HSE)[75] for an accurate treatment of the valence band, conduction band, and lithium defect levels. Use of the HSE functional has been shown to provide a large improvement in correcting the band gap underestimation prevalent in LDA/GGA DFT calculations[76] and also to provide more accurate defect level positions for a variety of insulating materials.[77] The fraction of Hartree-Fock exchange is fitted on a case-by-case basis to reproduce experimental band gaps for each material considered. Therefore, to obtain band gaps that agree with experimental results, Hartree-Fock exchange fractions of 0.45, 0.35, 0.40, 0.45, and 0.42 are used for $Al_2O_3$, $AlF_3$, $MgO$, $SiO_2$, and $ZrO_2$, respectively. In all cases, a 2x2x2 k-point mesh is utilized for HSE simulations and all other computational details are identical to the GGA calculations, including the supercell sizes. Li metal energies are recalculated with HSE using the appropriate exchange fractions for each compound. The HSE calculations are performed only for electronic structure and lithium formation energies and not for migration barriers due to the large computational cost of HSE migration barrier calculations. We believe that including the additional accuracy of HSE for the band gaps and alignments is essential to obtaining accurate results but that the impact of HSE on the values of the migration barriers compared to GGA is likely to be relatively small. In support of this assumption we note that Ref. [78] compared calculated migration barriers from HSE and LDA or GGA for a few systems and found energy differences of the order of only approximately 10%.

**Lithium insertion in crystalline and amorphous coatings**

For these calculations, we consider each coating material in its most stable crystalline form, as shown in Fig. 1. To understand how Li behaves in these coating materials, we must first know the charge state ($Li^0$ or $Li^+$) of Li in these materials. We determine this by calculating the formation energy of both $Li^0$ and $Li^+$ charge states as a function of the electron chemical potential (i.e., the Fermi energy). As a check we also consider the $Li^-$ charge state for GGA calculations only, however it was found not to be stable under all relevant conditions. The charged defect of lithium can be created when an electron is removed (for $Li^+$) from or added (for $Li^-$) to the $Li^0$-inserted supercell. Bader charge analysis is performed using codes developed by Henkelman *et al*.[79] and carried out in order to examine whether the electron of lithium is delocalized away from its nucleus. Since the distance between a Li and its nearest-neighbor images is about 10 Å in our supercells, we expect there to be only minor errors in the Li energies introduced by the finite size of the supercells. To verify this, we test MgO in detail. We find that the finite size effect error in the $Li^+$ formation energy for a 3 x 3 x 3 MgO supercell, which is typical for the size we use for all systems, is ~1% of that for an approximately infinite supercell, corresponding to 45 meV per $Li^+$ in MgO. The infinite supercell energy for $Li^+$ in MgO was estimated by calculating the formation energy of $Li^+$ with three supercells of sizes $L$ x $L$ x $L$ ($L$ = 2, 3, and 4), fitting to a linear function of $1/L$, and extrapolating to infinite $L$. We therefore consider our energies to be approximations to infinitely dilute Li in the cell.

The formation energy of lithium for a charge state $q$ ($q = 0$ for Li$^0$, $q = +1$ for Li$^+$, and $q = -1$ for Li$^{-1}$) can be written as[80-84]

$$\Delta E_f = E(\text{Li}^q\text{Host}) - E(\text{Host}) - \mu_{\text{Li}} + q(E_{\text{VBM}} + E_F + E_{\text{shift}}) \quad \text{(AI-1)}$$

and is a linear function of the Fermi energy, $E_F$. The Fermi level is given relative to the energy of the valence band maximum (VBM), $E_{\text{VBM}}$, of the perfect system. $E(\text{Host})$ and $E(\text{Li}^q\text{Host})$ are respectively the total energies of perfect (host) and Li$^q$-inserted (host+lithium defect) systems. $E_{\text{shift}}$ is the energy correction for the VBM of the charged defect system, determined by the average electrostatic potential energy.[80,81,85] The chemical potential of lithium $\mu_{\text{Li}}$ is obtained as the total energy of bulk lithium metal per lithium atom from our DFT calculation. This choice of reference means that the formation energy is referenced to lithium metal and therefore represents the energy of the reaction: Li + Host → Li$^q$Host + $q$e$^-$. Note that we do not consider possible reactions of Li with the coating compounds to form new phases, e.g., Li$_2$O. Such reactions could certainly influence the coating integrity and should be considered in future work, but are not the focus here. The present solubilities are therefore of relevance under conditions where transformation to new phases are kinetically inhibited, which could easily be the case at room temperature for many experimental operating conditions. We also determine the equilibrium concentration, $C$, of lithium and lithium ions, which is calculated assuming noninteracting Li by[86]

$$C = C_0 \times \int d(\Delta E_f) D(E) \frac{e^{-\beta(\Delta E_f)}}{1 + e^{-\beta(\Delta E_f)}} \approx C_i \frac{e^{-\beta(\Delta E_{f,i})}}{1 + e^{-\beta(\Delta E_{f,i})}} \quad \text{(AI-2)}$$

where $C_i$ is the concentration of interstitial sites per unit volume of type $i$, $\Delta E_{f,i}$ is the defect formation energy for a defect of type $i$, $D(E)$ is the density of Li interstitial states per unit energy normalized to one (note that this Li interstitial density of states should not be confused with the electronic density of states which is often denoted with a similar symbol), $C_0$ is the average number of Li sites per cm$^3$ that can be simultaneously occupied, and $\beta = 1/k_BT$, where $k_B$ is the Boltzmann constant and the $T$ is the temperature ($T$=300 K). The first equality will be used for amorphous materials, which have a distribution of site energies given by $D(E)$. For the amorphous materials we take $C_0 = 8.11 \times 10^{22}$ cm$^{-3}$, which is the Li concentration of Li$_2$O, and provides a reasonable upper bound to number of available Li sites. Note that the total concentration of available sites in the amorphous system can be easily calculated from $D(E)$, but as it is not clear how many of these sites can be simultaneously occupied it is unclear how to use this value to calculate $C_0$. We therefore use the estimated value from Li$_2$O instead. The approximate equality at the end of Eq. (AI-2) applies when the system is dominated by interstitial sites of type $i$ with a single formation energy, $\Delta E_{f,i}$, and in this case $C_i$ is the concentration of interstitial sites per unit volume of type $i$. Eq. (AI-2) is technically only correct for non-interacting Li but we use it as an approximate guide in these calculations.

To estimate the lithium solubilities from the distributions of formation energies reported in Ref. [37], which was calculated for Li$^0$ only, we use Eq. (AI-2). The $D(E)$ term can be approximated by a Gaussian function fit to the formation energy data reported in Ref. [37], which energies have a mean and standard deviation of 0.55 eV and 0.50 eV for am-Al$_2$O$_3$ and 0.68 eV and 0.40 eV for am-AlF$_3$, respectively. Note that the standard deviations can easily be larger than the mean as negative formation energies are included in the distribution. Using these $D(E)$ and Eq. (AI-2) the values of the Li solubility at $T$ = 300 K for am-Al$_2$O$_3$ and am-AlF$_3$, are approximately $C = 1.2 \times 10^{22}$ cm$^{-3}$ and $C = 3.32 \times 10^{21}$ cm$^{-3}$, respectively (see Table AI-2). Note that Eq. (AI-2) assumes that the Li are non-interacting. This assumption will certainly not hold rigorously at the high Li concentration in am-Al$_2$O$_3$ and am-AlF$_3$. However, entropy effects are small at room temperature and unless the Li interact very strongly the large number of negative formation energy states in the $D(E)$ distribution in Ref. [37] assure that a high Li concentration will be obtained. In fact, a direct calculation by Jung, et al,[38] of the number of stable Li in am-Al$_2$O$_3$ strongly supports our estimate, as discussed further below. As discussed above, Li$^0$ ionizes to Li$^+$ and an electron in the conduction band, which means that the formation energy for Li$^0$ is an upper bound for the formation energy for Li$^+$, with both energies being equal in the dilute Li limit for a strongly n-type material. Therefore, for any Fermi level less than or equal to the CBM the Li$^+$ solubility will be greater than or equal to that calculated above for Li$^0$. We therefore consider the solubility for the Li$^+$ ion at $T$ = 300 K as $C \geq 1.2 \times 10^{22}$ cm$^{-3}$ for am-Al$_2$O$_3$ and $C \geq 3.32 \times 10^{21}$ cm$^{-3}$ for am-AlF$_3$. We stress that these values are quite approximate as they are all obtained from GGA formation energies. In particular, it has been shown here (see differences in defect formation energies between materials listed in Table AI-1 and Table AI-2) and in the literature that HSE defect formation energies can differ on the order of 1 eV from GGA or LDA calculations,[87] and in addition there is significant support that the HSE values are more accurate (e.g., HSE produces an interstitial insertion energy for H in ZnO close to experimentally measured values[87,88]) Generally, the insertion energy of neutral Li is higher for HSE than GGA, and the insertion energy of Li$^+$ in the n-type limit ($E_{Fermi}$ at CBM) is also higher for HSE than GGA, except for crystalline AlF$_3$. Consistent with this trend, a direct calculation of Li insertion into one site in am-AlF$_3$ with GGA and HSE resulted in the energy to insert Li being 0.24 eV higher for HSE over GGA. Therefore, we suggest that GGA provides an upper bound for Li solubilities in these materials by virtue of their lower formation energies. For am-Al$_2$O$_3$ and am-AlF$_3$, even though the Li intercalation energy may need to be shifted up by about 1 eV to account for an approximate GGA to HSE formation energy shift, the solubility change will be relatively small. Because of the spread of the values of $D(E)$ in Eq. (AI-2), even if the formation energy is shifted up by 1 eV, there are still many states with low energy that dominate intercalation. Based on Eq. (AI-2) we can calculate that if a new $D(E)$ was shifted up by 1 eV to have a mean and standard deviation, respectively, of 1.55 (=0.55+1) eV and 0.50 eV for am-Al$_2$O$_3$, the new $C(Li)$ is only scaled by a factor of 0.007 ~ 10$^{-2}$ compared with our original $C(Li)$ value. This two orders of magnitude decrease would not have a significant qualitative effect on our following calculations and discussion for the amorphous coating materials. We note that it is sometimes convenient to estimate a single "effective" Li formation energy ($\Delta E_f^{eff}$) for am-Al$_2$O$_3$ and am-AlF$_3$ corresponding to their estimated Li solubilities obtained using Eq. (AI-2). Such a value can then be compared quickly to systems with just one formation energy. The simplest approach is by fitting $\Delta E_f^{eff} = \Delta E_{f,i}$ so that the second approximate equality in Eq. (AI-2) holds, taking $C_i = C_0$ in Eq. (AI-2). This approach yields $\Delta E_f^{eff} = 0.049$ eV and $\Delta E_f^{eff} = 0.082$ eV for am-Al$_2$O$_3$ and am-AlF$_3$, respectively (see Table AI-2).

**Results of lithium interstitial defect formation**

Figure AI-1 presents the total density of states (DOS) profiles using HSE calculations for perfect and Li$^0$- and Li$^+$- inserted systems for all coating materials studied in this work. Some general features of the electronic behavior of interstitial lithium inserted in these coating materials can be seen clearly in Fig. AI-1. First, there is no lithium defect state (Li$^0$/Li$^+$ energy level) occurring in the band gap. Second, the valence electron of lithium in the Li$^0$-inserted system goes directly to the conduction band and moves the Fermi level (for DFT calculations, Fermi level is the energy of the highest occupied state at $T = 0$ K) to the bottom of conduction band. In contrast, the Fermi levels of the perfect and Li$^+$- inserted systems are at the top of valence band, as would be expected. Note that for Al$_2$O$_3$, MgO and to a lesser extent AlF$_3$ that the Fermi level for Li$^0$ is above the CBM. This positioning is due to the low DOS at the CBM for these materials, and the finite Li concentration in the supercells used for these calculations. In the limit of the infinite supercell size and infinitely dilute Li, it is expected that the Fermi level for Li$^0$ will become equal to the CBM.

Examination of the defect formation energetics of Li in various charge states for each insulating coating is useful for further understanding the electronic behavior of interstitial Li in each coating material, and provides a direct means for calculating the solubility of Li in its stable charge state for each coating material by using Eq. (AI-2). The plots of HSE formation energies versus Fermi energy are shown in Fig. AI-2, where the value of $q$ is the slope of each line. It is shown that for the whole range of Fermi energy level variation (from the VBM to CBM for each material), the formation energy of Li$^+$ ion is lowest and there is no transition of charge states below the CBM. This result implies that the Li$^+$ ion is the only stable charge state when lithium dissolves into the coating materials, which is consistent with the fact that lithium spontaneously ionizes in these coating materials and will be in the form of a Li$^+$ ion. Our Bader charge analysis also confirms this ionization behavior by showing that the valence electron of lithium is delocalized away from its nucleus in the case of Li$^0$ insertion. The charge (negative) left on the Li nucleus is about 0.06~0.25 in the five crystalline cases we calculated, consistent with significant electron delocalization. The HSE formation energies plotted in Fig. AI-2 are tabulated in Table AI-1. As a reference, we have also included the calculated GGA formation energies in Table AI-2. The large changes, often more than one eV, show the impact of using the HSE approach in place of GGA for the Li formation energies. For both Tables AI-1 and AI-2, the Li interstitial concentrations for Li$^0$ insertion (i.e., from an uncharged DFT calculation where the Li electron is allowed to go where it wants in the coating) and Li$^+$ insertion (i.e., from a charged DFT calculation with a removed electron, which is placed at the VBM or CBM of the coating) were calculated using Eq. (AI-2). We note again that the differences between the Li$^0$ energy and concentration and the Li$^+$ energy and concentration for the electron placed at the CBM come from the finite size cell used in the Li$^0$ calculations and would be expected to go to approximately zero in the limit of very dilute Li. The Li solubility for amorphous Al$_2$O$_3$ (am-Al$_2$O$_3$) and AlF$_3$ (am-AlF$_3$) materials are also given for comparison with their crystalline counterparts. Table AI-3 tabulates the experimental and calculated band gaps between GGA and HSE calculations for all coatings. The band gap shift is simply calculated as the difference between HSE and GGA band gaps. We note that the amorphous system energies are all taken from GGA calculations, which were detailed in Ref. [37].

To validate the predictions of Li preferring to ionize to Li$^+$ in the materials considered here, we examine the band alignment of the coating materials considered in this work with respect to the Li/Li$^+$ redox level. The predicted Li ionization suggests that the CBM level for each insulating coating considered in this work should approximately lie below the Li/Li$^+$ redox level. In this way, the CBM levels are lower in energy than the energy to make neutral Li from Li$^+$, and Li will thus prefer to ionize when entering the coating materials. Fig. AI-3 shows the literature values for experimentally determined band alignment of various semiconductors and insulators (as well as the insulating coatings considered in this work) versus important energy references such as the vacuum level, standard hydrogen electrode (SHE), and the Li/Li$^+$ (aqueous solution) level.[89-92] For all five of the materials here the Li/Li$^+$ level is within 1 eV of the CBM, with Li/Li$^+$ slightly higher for ZrO$_2$ and MgO and somewhat lower for SiO$_2$, Al$_2$O$_3$ and AlF$_3$. The absence of any Li/Li$^+$ states in the band gap for the latter three materials suggests either that the alignment of the insulators and Li/Li$^+$ energies in HSE has some errors or that there is a destabilization of the energy difference $\Delta E = E(\text{Li}^+) - E(\text{Li})$ for Li in these insulators as compared to $\Delta E$ for aqueous Li$^+$ and metallic Li, which would not be unexpected. Whatever the explanation for these relatively small differences, the band diagram is qualitatively consistent with the calculations in that the Li/Li$^+$ level is quite close to the CBM for all these materials. We also note that the band diagram suggests that a typical cathode would not oxidize or reduce the insulating oxide coatings considered in this work. While this implication is correct in terms of simple electron flow between the structures, the band alignment should not be taken to imply that the coatings are stable with respect to typical cathodes, as this simple band picture does not represent all the energetics of a possible chemical reaction between a cathode and coating material.

**Table A1-1.** Formation energy and the equilibrium concentration (from Eq. (AI-2)) at 300 K in crystalline coating materials from HSE calculations. $E_F$ is the Fermi energy measure relative to the valence band maximum.

| Material | Formation energy vs Li/Li$^+$ $\Delta E_f$ (eV/Li) | | | Equilibrium concentration (i.e., solubility) at room temperature ($T$=300K) ((# Li/cm$^{-3}$) | | |
|---|---|---|---|---|---|---|
| | Li ($E_F = E_g$) | Li$^+$ | | Li ($E_F = E_g$) | Li$^+$ | |
| | | $E_F = 0$ | $E_F = E_g$ | | $E_F = 0$ | $E_F = E_g$ |
| Al$_2$O$_3$ | 4.33 | -6.63 | 2.25 | 2.42x10$^{-51}$ | 1.12x10$^{22}$ | 1.97x10$^{-16}$ |
| AlF$_3$ | 3.14 | -8.12 | 2.59 | 3.43x10$^{-31}$ | 2.14x10$^{22}$ | 8.17x10$^{-22}$ |
| MgO | 6.31 | -2.05 | 5.33 | 1.28x10$^{-83}$ | 1.05x10$^{23}$ | 2.87x10$^{-67}$ |
| ZrO$_2$ | 5.90 | 0.00 | 5.82 | 2.54x10$^{-77}$ | 1.49x10$^{22}$ | 5.94x10$^{-76}$ |
| SiO$_2$ | 2.45 | -6.70 | 2.27 | 2.79x10$^{-19}$ | 3.28x10$^{22}$ | 2.69x10$^{-16}$ |

**Table AI-2.** Formation energy in crystalline coating materials, effective formation energy for amorphous materials and the equilibrium concentration (from Eq. (AI-2)) at 300 K from GGA calculations. $E_F$ is the Fermi energy relative to the valence band maximum.

| Material | Formation energy vs Li/Li$^+$ (eV/Li) | | | Equilibrium concentration (i.e., solubility) at room temperature ($T$=300K) (# Li/cm$^{-3}$) | | |
| --- | --- | --- | --- | --- | --- | --- |
| | Li ($E_F = E_g$) | Li$^+$ | | Li ($E_F = E_g$) | Li$^+$ | |
| | | $E_F = 0$ | $E_F = E_g$ | | $E_F = 0$ | $E_F = E_g$ |
| $Al_2O_3$ | 3.56 | -2.56 | 2.70 | $1.87 \times 10^{-38}$ | $1.12 \times 10^{22}$ | $4.36 \times 10^{-24}$ |
| $AlF_3$ | 1.95 | -5.59 | 1.48 | $4.17 \times 10^{-11}$ | $2.14 \times 10^{22}$ | $2.77 \times 10^{-03}$ |
| MgO | 4.89 | 0.18 | 4.37 | $7.82 \times 10^{-60}$ | $1.12 \times 10^{20}$ | $4.84 \times 10^{-51}$ |
| $ZrO_2$ | 4.72 | 1.56 | 4.57 | $1.62 \times 10^{-57}$ | $1.82 \times 10^{-4}$ | $5.12 \times 10^{-55}$ |
| $SiO_2$ | 1.26 | -4.27 | 0.98 | $1.8 \times 10^{1}$ | $3.28 \times 10^{22}$ | $1.38 \times 10^{6}$ |
| am-$Al_2O_3$[37] | 0.049 | | ≤ 0.049 | $1.2 \times 10^{22}$ | | ≥ $1.2 \times 10^{22}$ |
| am-$AlF_3$[37] | 0.082 | | ≤ 0.082 | $3.32 \times 10^{21}$ | | ≥ $3.32 \times 10^{21}$ |
| am-$Li_{3.5}Al_2O_3$[38] | | | | $3.03 \times 10^{22}$ | | |

**Table AI-3.** Values of band gap of the materials obtained from our HSE and GGA calculations, along with a comparison to experimental band gap values. Simply subtracting the HSE and GGA energy gaps yields the gap change. All values are given in eV.

| Material | GGA Calculated $E_g$ (eV) | HSE Calculated $E_g$ (eV) | Experiment $E_g$ (eV) | Gap change |
|---|---|---|---|---|
| $Al_2O_3$ | 5.26 | 8.88 | 8.8 [Ref. [91]] | 3.62 |
| $AlF_3$ | 7.07 | 10.71 | 10.8 [Ref. [89]] | 3.64 |
| MgO | 4.19 | 7.38 | 7.5 [Ref. [91]] | 3.19 |
| $ZrO_2$ | 3.01 | 5.82 | 5.8 [Ref. [91]] | 2.81 |
| $SiO_2$ | 5.25 | 8.97 | 9.0 [Ref. [91]] | 3.72 |

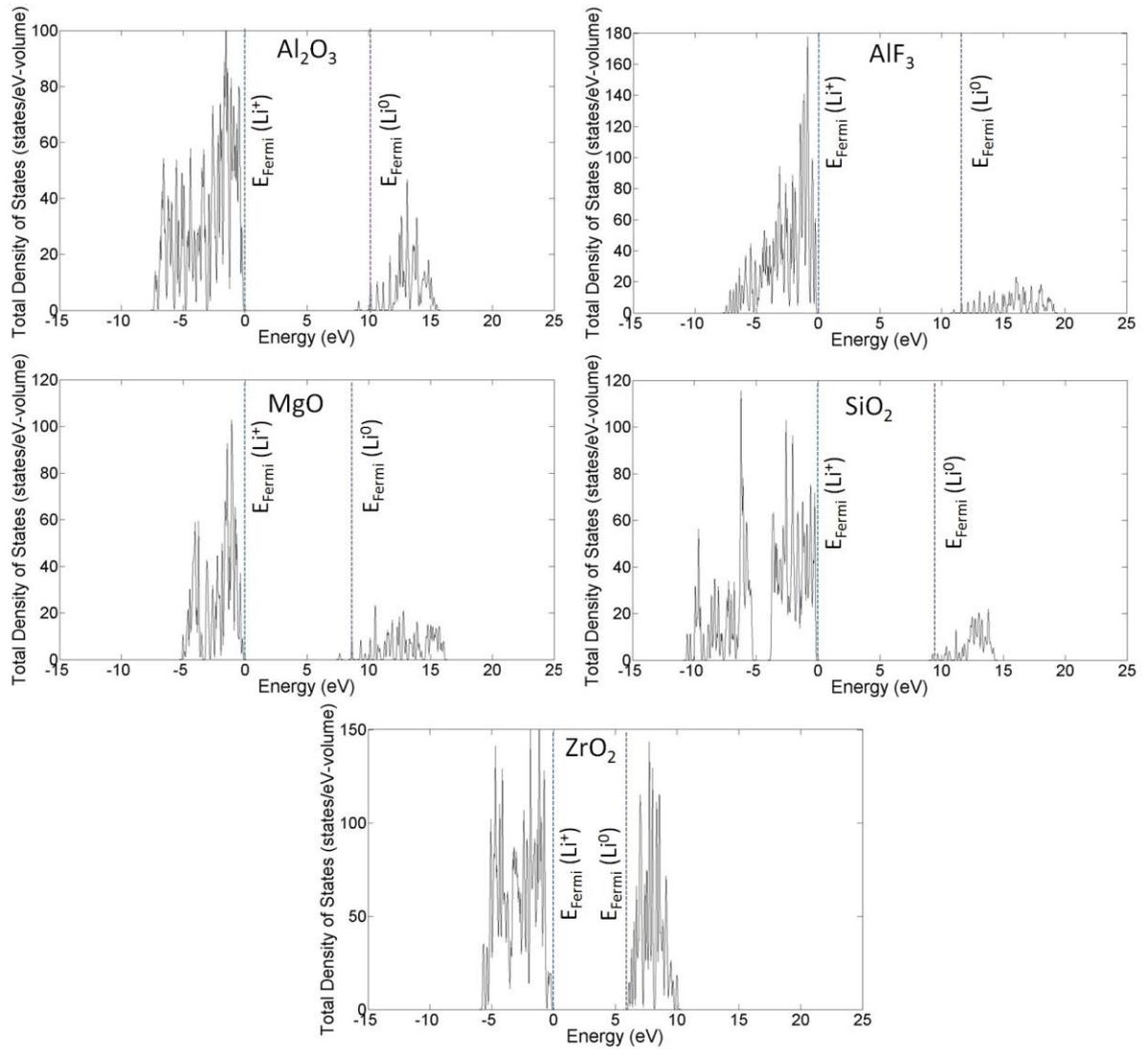

**Fig. AI-1.** Total densities of states (DOS) for each coating material using HSE. The vertical dotted lines indicate the position of the Fermi energy for the $Li^0$ (blue dotted line near the VBM) and $Li^+$ (purple dotted line near the CBM) cases. The plots shown here are for the pristine coating materials, and the Fermi energies from calculations of Li and $Li^+$ interstitial formation have been added onto these DOS plots. Note that for $Al_2O_3$, MgO and to a lesser extent $AlF_3$ that the Fermi level for $Li^0$ is above the CBM. This is due to the low DOS at the CBM for these materials, and the finite Li concentration in the supercells used for these calculations. In the limit of infinite supercell size, it is expected that the Fermi level for $Li^0$ proceeds to the CBM.

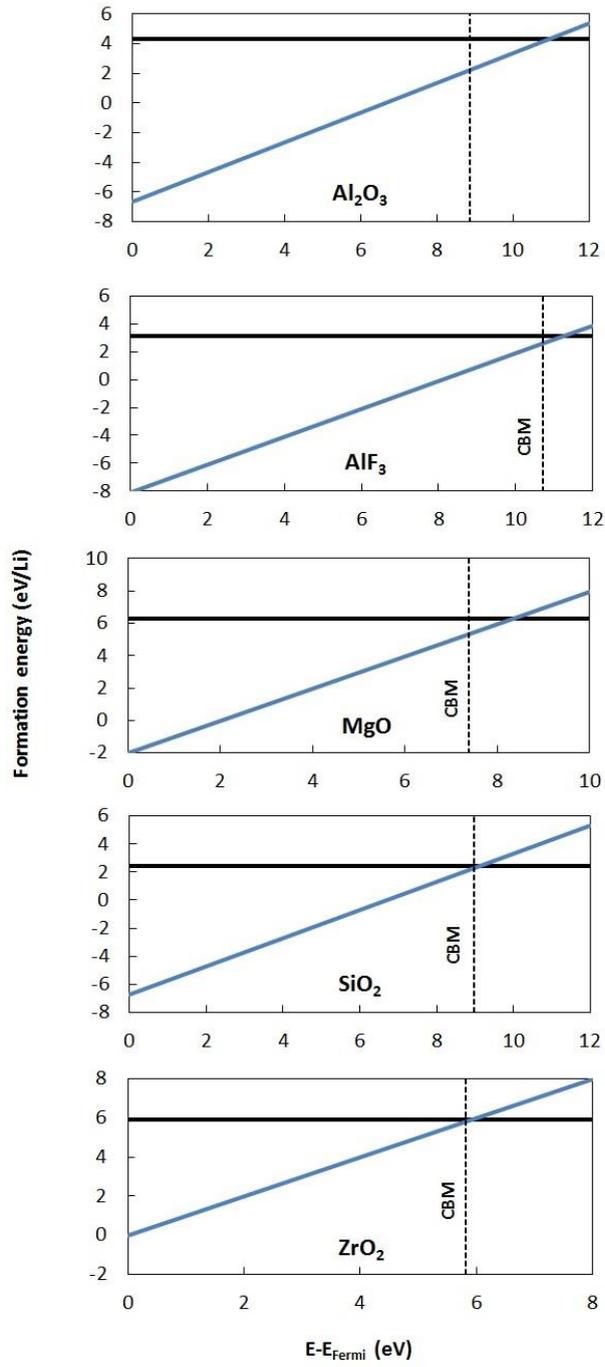

**Fig. AI-2.** Plots of HSE formation energies for all coating materials as a function of Fermi energy. Two charge states of Li dopant are considered, namely, $Li^0$ (neutral $q=0$, flat solid black line) and $Li^+$ (positive $q=+1$, sloped solid blue line), and the $q$ values are the slopes of solid lines. The black vertical dashed line indicates the position of the CBM. The zero of energy is taken as the VBM in all cases. Note that the crossing point of $Li^0/Li^+$ is above the CBM in all cases.

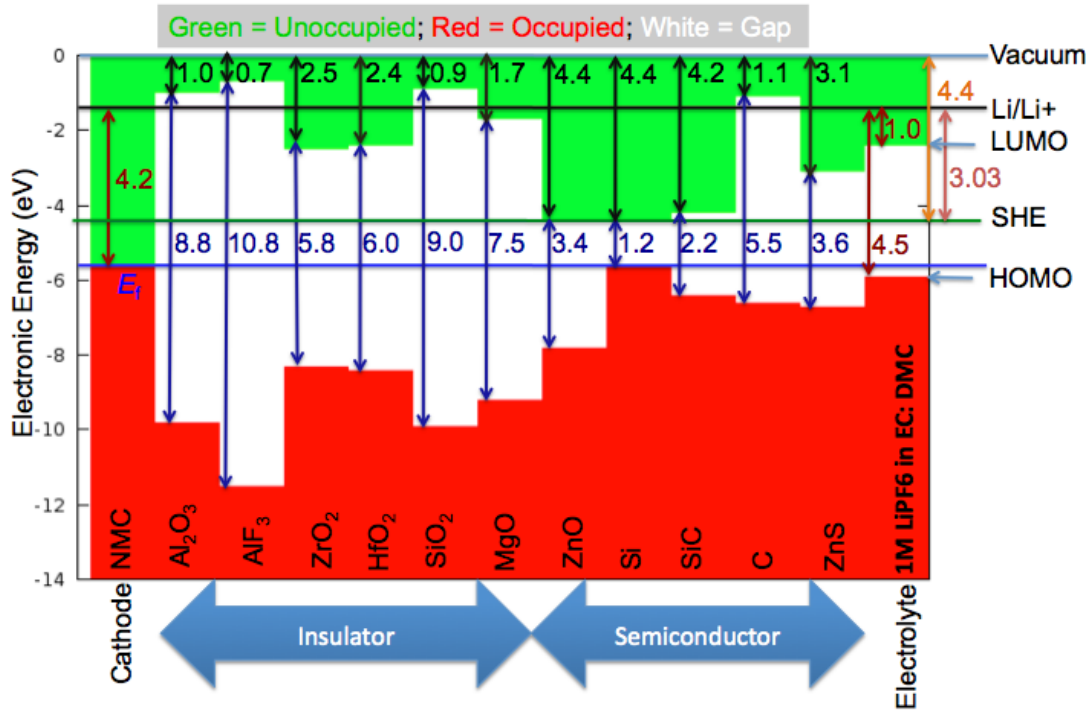

**Fig. AI-3.** Absolute band alignment of conduction and valence band edges for various insulators and semiconductors. The relevant reference levels are included: vacuum energy, standard hydrogen electrode (SHE), and Li/Li$^+$ in aqueous solution. Typical cathode and electrolyte material are included for comparison. NMC is a typical LiNi$_{1/3}$Mn$_{1/3}$Co$_{1/3}$O$_2$ cathode material. The highest occupied molecular orbital (HOMO) and lowest unoccupied molecular orbital (LUMO) levels of 1M LiPF$_6$ in EC:DMC as a typical liquid electrolyte are provided. Note that the Li/Li$^+$ level is above the CBM for MgO and ZrO$_2$ but lies within the band gap of Al$_2$O$_3$, AlF$_3$ and SiO$_2$. Experimental data used in this band alignment were collected from Ref. [89] for AlF$_3$, Ref. [90] for Al$_2$O$_3$, ZrO$_2$, HfO$_2$, SiO$_2$, MgO, and ZnO, Ref. [91] for Si, SiC, C (diamond), and ZnS, and Ref. [92] for the electrolyte.

**Appendix II: Detailed Structures and Li Migration Pathways in Crystalline Phases**

This appendix gives a detailed discussion of structures and Li migration pathways and energetics in each of the insulating coating materials considered in the present work.

**α-AlF$_3$**

Aluminum fluoride has several polymorphs,[93] of which α-AlF$_3$ (α-phase) is thermodynamically the most stable at room temperature. It has a rhombohedral structure with space group $R\bar{3}c$ and is transformed to the cubic phase of ReO$_3$ perovskite-type structure with space group $Pm3m$ upon heating above ~730 K.[93-95] By exploring several trial interstitial sites, our DFT calculations show that there is a stable interstitial site located at the center of each distorted cube of α-AlF$_3$, shown as lithium atoms labeled with numbers 1 and 3 in Fig. AII-1(a). It is also shown in Fig. 1(b) that there are two equivalent interstitial sites per primitive cell respectively located at $\left(\frac{1}{4},\frac{1}{4},\frac{1}{4}\right)$ and $\left(\frac{3}{4},\frac{3}{4},\frac{3}{4}\right)$ in direct coordinates. The interstitial lithium is surrounded immediately by 8 oxygen atoms of the first shell with the Li-O distance $d_{Li-O}$ = 2.53 Å, then by 8 aluminum atoms of the second shell with Li-Al distance $d_{Li-Al}$ = 3.14 Å. Lithium ions can diffuse by hopping from one interstitial site to its nearest neighbors, e.g., between point 1 and point 3 shown in Fig. AII-1(a). The hop has a distance of 3.59 Å and needs to surmount an energy barrier of $E_m$ = 0.929 eV, as shown in the migration energetic profile in Fig. AII-1(d). The transition state of the hop is the state where the lithium ion climbs up at the saddle point denoted as point 2 in Fig. AII-1(a). The saddle point is surrounded with 4 nearest-neighbor oxygen atoms of distance 1.58 Å, and with 4 next-nearest-neighbor aluminum atoms of distance 2.52 Å.

**α-Al$_2$O$_3$**

α-Al$_2$O$_3$ is the commonly-occurring and stable crystalline polymorphic phase of aluminum oxide (alumina). Its hexagonal corundum structure (space group $R\bar{3}c$) is shown in Fig. AII-1(b). Aluminum and oxygen atoms occupy the 12$c$ and 18$e$ sites, respectively.[96] There are 6$b$ interstitial sites occupying the middle point of alternate pairs of aluminum atoms. The α-Al$_2$O$_3$ structure can be represented by stacking close-packed oxygen atom layers with interweaved aluminum atoms along the <001> direction. Aluminum atoms are the centers of edge-sharing octahedra and occupy a hexagonal lattice having one third of the sites unoccupied, which are designated as octahedral interstitials. The primitive cell belongs to the trigonal crystal system.

There is one interstitial site per primitive cell at $\left(\frac{1}{2},\frac{1}{2},\frac{1}{2}\right)$ as shown in Fig. 1(a). Each interstitial lithium is surrounded with 6 oxygen and 2 aluminum atoms of distance $d_{Li-O}$ ≈ 1.96 Å, $d_{Li-Al}$ ≈ 2.17 Å. It is shown from our CI-NEB calculations that, in order to hop a distance 3.5 Å between two nearest-neighbor interstitial sites (point 1 and point 3 in Fig. AII-1(b)), lithium needs to transit the saddle point (point 2) and overcome an energy barrier of $E_m$ = 2.498 eV as shown in Fig. AII-1(d).

**c-MgO**

There are three crystal phases of magnesium oxide (MgO), namely, $B_1$ (NaCl), $B_2$ (CsCl), and $B8_1$ (inverse NiAs).[97] $B_1$ (denoted as c-MgO and the structure of interest in this work) has a rocksalt cubic structure of NaCl (space group $Fm\bar{3}m$), which is the stable phase at room temperature and pressure conditions. In the rocksalt structure, magnesium atoms have a face-centered cubic (fcc) arrangement, with oxygen atoms occupying all the octahedral holes. Equivalently, this structure can be described as an fcc lattice of oxygen atoms with magnesium atoms in the octahedral holes. Each type of atoms has a coordination number of 6.

The stable interstitial site is located at the center of each smallest cube made by oxygen and magnesium atoms (e.g., at point 1 or 3 shown in Fig. AII-1(c)). Each primitive cell has two interstitial sites at $\left(\frac{1}{4},\frac{1}{4},\frac{1}{4}\right)$ and $\left(\frac{3}{4},\frac{3}{4},\frac{3}{4}\right)$ in direct coordinates as shown in Fig. 1(c). Consequently, each interstitial lithium ion is surrounded immediately by 4 Mg$^{2+}$ cations and 4 O$^{2-}$ anions, with $d_{Mg-Li} = d_{O-Li}$ = 1.845 Å in the ideal structure. Under the cubic symmetry of rocksalt structure, the lithium ion may migrate three-dimensionally in the c-MgO crystal by hopping a distance 2.13 Å from one interstitial site to one of its six nearest-neighbor interstitial sites with equal probabilities. One of the equivalent migration pathways, path 1-2-3, is shown in Fig. AII-1(c), where points 1, 2, and 3 are the initial, the saddle, and the end points, respectively. According to our CI-NEB calculations, the lithium ion hopping energy barrier is $E_m$ =1.419 eV, as shown in Fig. AII-1(d).

**m-ZrO$_2$**

Zirconium dioxide (zirconia), $ZrO_2$, occurs in nature mostly in the mineral baddeleyite and has the monoclinic crystal structure with space group $P2_1/c$ (denoted as m-$ZrO_2$ and the structure of our interest in this work) at room temperature, which transforms to tetragonal (space group $P4_2/nmc$) and cubic (space group $Fm\overline{3}m$) phases at high temperatures. There are 4 $ZrO_2$ in the unit cell as shown in Fig. 1(d). In addition, the structure possesses two interesting features.[98] Firstly, zirconium has a sevenfold coordination with oxygen atoms and the nearest Zr-O distance varies from 2.04 Å to 2.26 Å and the next-nearest Zr-O distance is 3.77 Å. Secondly, there is an interesting alternation of fluorite-like layers parallel to <100>, one contains oxygen atoms in triangular coordination and the other contains oxygen atoms in tetrahedral coordination, which accounts for the strong tendency to twin on <100>.[98]

Our DFT calculations suggest that there exist two pairs of stable interstitial sites per unit cell, which are symmetric about (0, 0, 0) as shown in Fig. 1(d). One pair is at (0.12, 0.55, 0.05) and (0.10, 0.93, 0.55) while the other (0.88, 0.45, 0.95) and (0.90, 0.07, 0.45) in direct coordinates. Our finding of the interstitial sites in m-$ZrO_2$ is consistent with that reported by Jiang *et. al*.[99] If one repeats the unit cell periodically in space, one could see that the interstitial sites will form a regular structure of double layers 5.22 Å apart and parallel to the <001> plane as shown in Figs. AII-2(a) and AII-2(b). In each layer, the connection between nearest-neighbor interstitials creates crystallographically equivalent zig-zag pathways running along <001> (<001> zig-zag pathways) such as paths 1-2-3-4 and 5-6-7-8 in one plane and paths 1'-2'-3'-4' and 5'-6'-7'-8' in another plane shown in Fig. AII-2(a). These lithium ion migration pathways in the m-$ZrO_2$ crystal have the lowest energy barrier of $E_m$=0.969 eV as shown in Fig. AII-2(c). Lithium ion migration along <010> by hopping between adjacent <001> zig-zag pathways, e.g., path 2-7 in Fig. AII-2, faces a higher energy barrier of $E_m$=1.101 eV. Finally, migration along <100>, e.g., path 3-3'-3-3' (i.e., a long 3-3' hop, then a short 3'-3 hop, and one more long 3-3' hop shown in Fig. AII-2(b)), is found unlikely because the energy barrier for the long 3-3' hop is very high, with $E_m$=3.642 eV (although the energy barrier for the short 3'-3 hop is as low as 0.05 eV). Therefore, these results suggest that $Li^+$ ion diffusion in m-$ZrO_2$ crystal is effectively one-dimensional along <001>.

**α-SiO$_2$**

Silicon dioxide (Silica), $SiO_2$, is well-known oxide used in a variety of applications. In nature, it is commonly found as sands or quartz, and has the stable form of α-quartz with space group $P3_121$ under room temperature and pressure conditions. The unit cell of α-quartz is shown in Fig. 1(e) and Fig. AII-3(a). Silicon atoms in α-quartz tetrahedrally coordinate with surrounding 4 oxygen atoms to create a ring-like network of vertex-sharing $SiO_4$ tetrahedra, yielding the net chemical formula $SiO_2$, and there are 3 $SiO_2$ per unit cell.

Our DFT calculations show that there are four possible stable interstitial sites per unit cell locating along the *c*-axis at (0, 0, 0), (0, 0, 0.34), (0, 0, 0.5) and (0, 0, 0.66), which are labeled with numbers 1 to 4 as shown in Fig. AII-3(a). Due to there being a fairly complex energy landscape with four sites, we initially explore the diffusion using first principles molecular dynamics (FPMD) simulations (not shown). The FPMD simulations are carried out at 1200 K, 1500 K, and 1700 K in a time period of 20 ps and we find one-dimensional diffusion along the <001> direction. We therefore focus our CI-NEB calculations along this path. We additionally explore selected hops along <100> and <010>. In total, we consider the migration pathways: 1-7, 3-6, and 1-2-3-4-5, which are respectively along <100>, <110>, and <001>. The energetic profiles for these migration pathways are shown in Fig. AII-3(b). Migration along <110> is found to have the highest energy barrier with $E_m$=1.064 eV. The <100> pathway has an energy barrier of $E_m$=0.736 eV. In contrast, migration along <001> has a substantially low energy barrier of $E_m$=0.276 eV. Therefore, $Li^+$ ion migration in α-quartz is expected to be primarily one-dimensional along the *c* axis. We will discuss in Sec. IIIC how this fast diffusivity of α-quartz makes it stand out as having potentially having fast enough transport for conformal cathode coatings.

Before ending this subsection, it is worth noting that $Li^+$ ion diffusion in α-quartz has been a subject of intensive investigations both experimentally and theoretically for decades. Verhoogen[54] first carried out a study of ionic diffusion and electric conductivity in natural α-quartz crystals and also observed that diffusion was one-dimensional. Theories have been developed to understand ionic transport in these materials,[55] majorly focused on $Al^{3+}$ cations substituting for $Si^{4+}$ with commonly charge-compensated by interstitial monovalent cations such as $Na^+$, $Li^+$, and $H^+$, or the holes that locate nearby the $Al^{3+}$ ions to form Al-Li, Al-Na, Al-OH, Al-hole centers. Obviously, $Al^{3+}$ defect plays a crucial role on the diffusion of $Li^+$ or $Na^+$ in these α-quartz crystals, which is not the case in our problem above of the pure α-quartz crystal without $Al^{3+}$ defect. However, we would like to stress here two relevant points: (i) It is generally accepted that ionic conductivity in quartz crystals is highly anisotropic ($\sigma_{//}/\sigma_\perp > 10^3$ where $\sigma_{//}$ and $\sigma_\perp$ are the values of the conductivity in directions parallel and perpendicular to the *c* axis (the z-optical axis), respectively).[55] Monovalent alkali ions move freely in open channels along the *c* axis. This trend is consistent to our DFT calculation above showing the *c* axis preferential diffusion of Li$^+$ ions; (ii) It is also generally accepted that the ionic conductivity of quartz crystals is attributed to the migration of alkali metal ions thermally dissociated from Al-M (or [AlO$_4$-M]$^0$) centers at temperature higher than 500 K, the dissociation reaction being [AlO$_4$-M]$^0$ ⇌ [AlO$_4$]$^-$ + [M]$^+$. Campone *et al.*[100] have reported for a synthetic quartz crystal that the dissociation and migration energies of the alkali-ions are 1.19 eV and 0.25 eV, respectively. This value of migration energy agrees very well with the lowest value of migration energy along the *c* axis of α-quartz from our DFT calculation in this work.

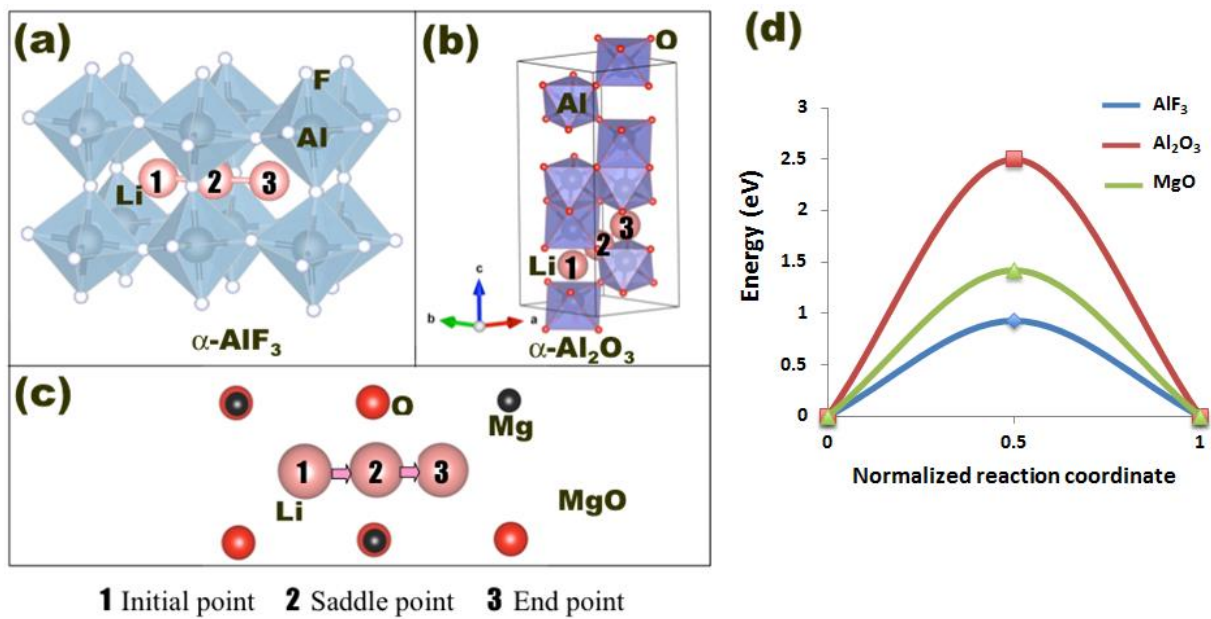

**Fig. AII-1** Interstitial lithium ion migration in α-AlF$_3$, α-Al$_2$O$_3$, MgO. Geometric presentation of the migration pathway in α-AlF$_3$ (a), α-Al$_2$O$_3$ (b), and MgO (c). CI-NEB energetic profiles for the migration pathways are presented in (d).

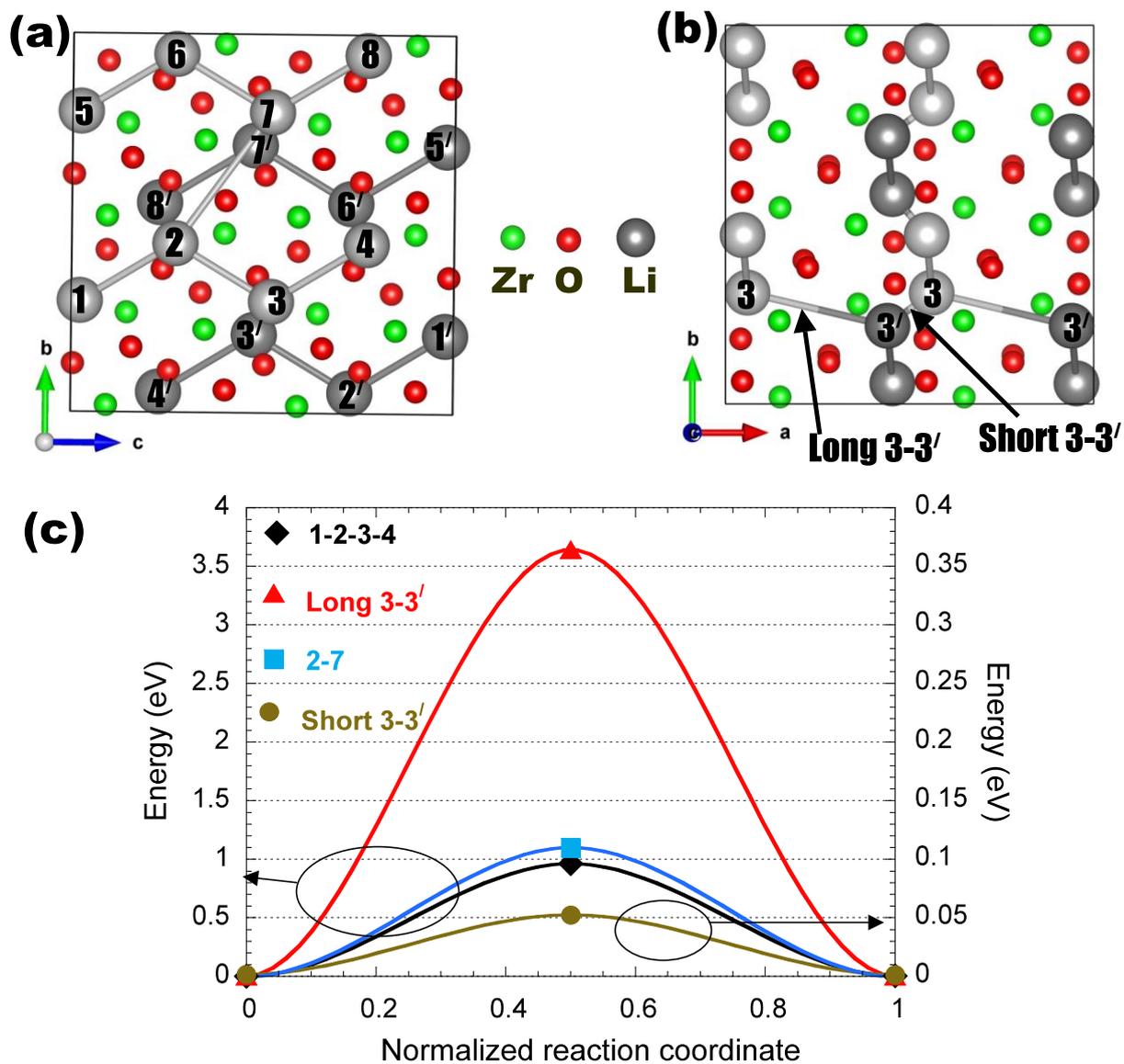

**Fig. AII-2** Interstitial lithium ion migration in m-ZrO$_2$. Geometric presentation of the possible migration pathways in m-ZrO$_2$ (a) and (b). CI-NEB energetic profiles for the migration pathways are presented in (c).

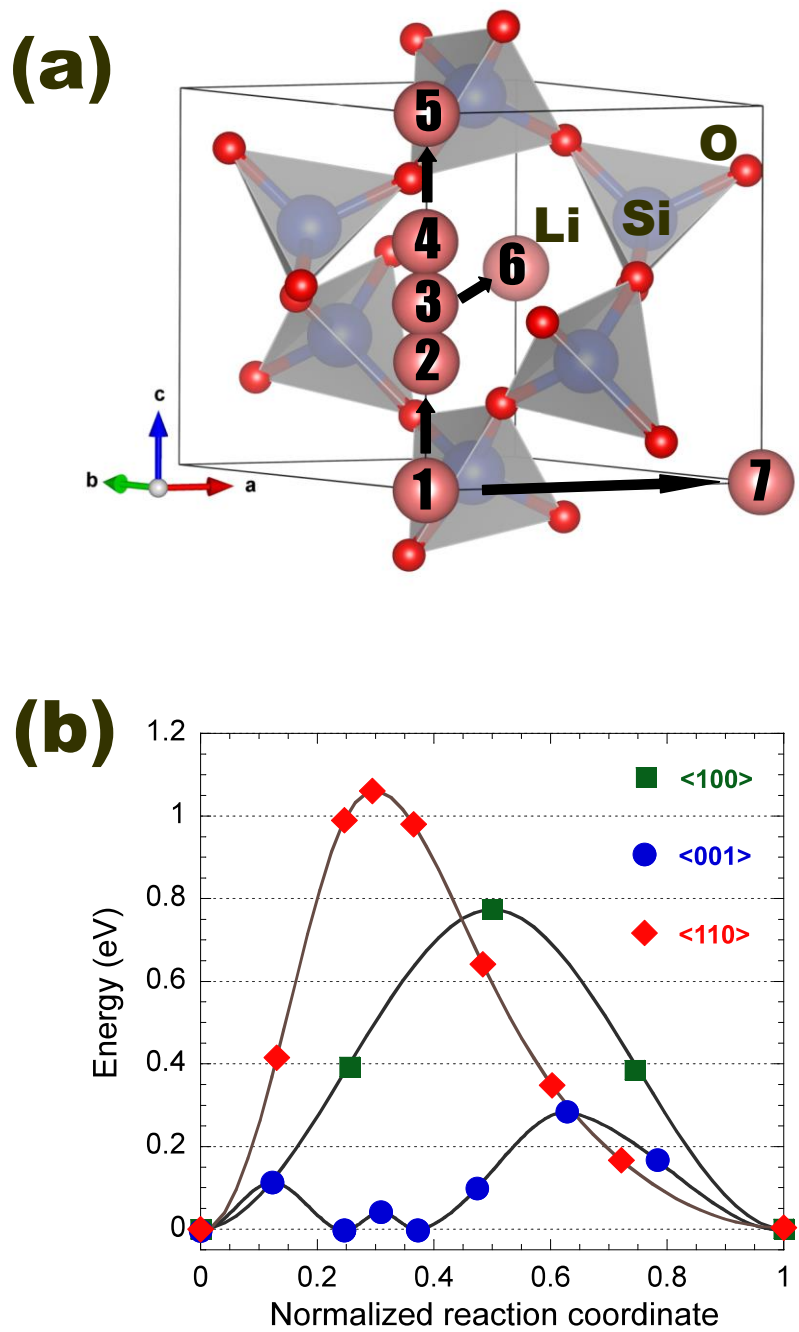

**Fig. AII-3.** Interstitial lithium ion migration in α-quartz SiO$_2$. Geometric presentation of the possible migration pathways in α-quartz (a). CI-NEB energetic profiles for the migration pathways are presented in (b).

**Appendix III: Estimates of Al$_2$O$_3$ Coating Resistivities from Previous Experiments**

This appendix describes how we estimate the resistivity of the am-Al$_2$O$_3$ coating from previous Atomic Layer Deposition (ALD) experiments. The coating is assumed to behave as an Ohmic resistor in series with an Ohmic resistance for the rest of the system. These assumptions yield the following expression for the voltage and current density relationship:

$$V = V_0 + R_{system}S_{system}J_{active} + \rho_{coating}L_{coating}J_{active} \qquad \text{(AIII-1)}$$

where $V$ is the applied voltage (or measured voltage) in the external circuit, $V_0$ is the open circuit voltage, $R_{system}$ is the total resistance of the system without the coating, $S_{system}$ is the effective area of the system (without the coating) that current flows through, $J_{active}$ is the current density through the coating layer, which is given per unit area of coating on the portion of the cathode surface active for Li-intercalation, $\rho_{coating}$ is the resistivity of the coating and $L_{coating}$ is the thickness of the coating. In the following analysis we obtain $V$, $J_{active}$ and $L_{coating}$ by fitting to published experimental measurements. We assume $R_{system}$, $S_{system}$ and $\rho_{coating}$ are the intrinsic properties of the system and they don't change with the $J_{active}$ and $L_{coating}$ variation for a given study. Starting from a set of $V$, $J_{active}$ and $L_{coating}$ values, linear fitting to $V$ as a function $J_{active}$ and $L_{coating}$ is performed to obtain the coefficients $V_0$, $R_{system}S_{system}$ and $\rho_{coating}$. The intercept of the fit determines $V_0$, and the coefficients of $J_{active}$ and $L_{coating}J_{active}$ provide the values of $R_{system}S_{system}$ and $\rho_{coating}$, respectively. For two of the references we fit (Ref. [47] and Ref. [51]), the charging/discharging rate for the experimental capacity-voltage plots was held fixed. Holding the charging/discharging rate fixed makes the $R_{system}S_{system}J_{active}$ term a constant for different coating thicknesses. Therefore, when we do the linear fitting for Ref. [47] and Ref. [51], we combine the $R_{system}S_{system}J_{active}$ term and $V_0$ term together and define $V_1 = V_0 + R_{system}S_{system}J_{active}$, and the intercept of the fit determines $V_1$. This $V_1$ is the open circuit voltage of the cathode plus the overpotential of the system resistance (without coating) and, for modest currents, should be qualitatively similar to the value of open circuit voltage of the cathode material.

To obtain the current density through the coating layer during the charging/discharging process we assume that the total current is uniformly flowing through an area of coating equal to the surface area of the cathode active for Li intercalation. We will assess both this active cathode surface area and its ratio to the geometric cathode disk surface area, as the latter is useful for determining current densities per unit active surface area from current densities given per unit geometric area. To help clarify these relationships we provide the relevant equations and definitions in Eq. (AIII-2):

$$J_{geom} = I/\text{cathode geometric surface area} = I/A_{geom}$$
$$J_{active} = I/\text{cathode surface area active for Li intercalation} = I/A_{active} \qquad \text{(AIII-2)}$$
$$J_{geom}/J_{active} = A_{active}/A_{geom} = b$$

Here, $J_{geom}$ is the current density per unit geometric surface area of the cathode, $J_{active}$ (also defined above) is the current density per unit surface area of the cathode active for Li intercalation, $A_{geom}$ is the geometric surface area of the cathode disk, $A_{active}$ is the surface area of the cathode active for Li intercalation, $I$ is the total current flowing through the battery, and $b$ is the unitless active surface area to geometric surface ratio. We wish to determine $J_{active}$ for a typical C-rate and also $b$ so we can easily transform $J_{geom}$ (which is often provided in papers) to $J_{active}$.

To determine $J_{active}$ we will consider 1 C rate current density through the coating, a value that can then be easily scaled to any C rate. To determine $b$ we will then compare our $J_{active}$ value with a previous experimentally reported 1 C rate $J_{geom}$ and use Eq. (AIII-2). For our analysis we use Li$_x$CoO$_2$ as the relevant cathode material because the experimental references that we used to fit resistivities choose either Li$_x$CoO$_2$ or Li$_x$(Ni$_{1/3}$Mn$_{1/3}$Co$_{1/3}$)O$_2$ (which we approximate as similar to Li$_x$CoO$_2$) in their work. We will assume the average radius of a primary cathode particle is ~1 μm for typical commercial Li$_x$CoO$_2$.[57,101] A 1 C rate means the charging/discharging time is 1 hour[49,101] and we will assume that this refers to Li$_x$CoO$_2$ being charged to about 140 mAh/g, or about 0.5 Li/Co, which is common practice to avoid phase changes and instability of the electrode.[102] We will assume that only the fraction of Li$_x$CoO$_2$ consisting of non-basal plane regions is active for Li transport, as Li cannot transport through the basal plane[103]. Based on previous DFT simulation work,[103] about ~50% of the total surface area is basal plane and 50% not basal plane. By considering a 1 μm particle with half its surface area active and capacity of 140 mAh/g we can use the density (5.05 g/cm$^3$) and molar mass (97.87 g/mol) of crystalline Li$_{x=1}$CoO$_2$ to estimate the current density through the active surface of the particle at a 1C rate to be about $J_{active}$=0.046 mA/cm$^2$. For each experiment below, we use either this value or closely related arguments to those just given to estimate $J_{active}$.

For Ref. [49] and Ref. [50] discussed below we will make use of the active surface area to geometric surface ratio $β$ to determine $J_{active}$. . Here we describe how the two surface areas and their ratio are determined. The geometric surface area of the tested cells is easy to obtain. Both references used CR-2032 cells. Although these two references didn't provide information on the cathode disk size, we obtained information about cathode disk size from other papers using similar CR-2032 coin cells. Specifically, other similar studies reported a geometric area of the cathode disk, $A_{geom}$, of 2 cm$^2$ [104], 1.6cm$^2$ [105], and 1.27cm$^2$ [29]. Here we choose the median number and set $A_{geom}$ = 1.6cm$^2$. To obtain the active surface area for the cathodes in Ref. [49] and Ref. [50], we can use the geometric arguments given above, but we need to know the amount of the active cathode material used in their experiments. However, the authors did not provide their active cathode material weight. We therefore estimate the weight they used based on typical cathode weights in commonly used protocols for CR-2032 coin cells. A literature search resulted in active cathode material weights for CR-2032 coin cell experiments of 5.1 mg[106], 5.14 mg[29], 5.8 mg[104], 19.7 mg[107] (this reference actually used a CR-2016 coil cell, but as the only difference is the thickness of the assembled cell we expect this to have a small impact on the total active material used), and 20 mg [108]. Based on these values from previous experiments, we find that the range of the cathode material weight is about 5~20mg. The average of the above five numbers is 11mg. We will use this number in the following estimation of □. Using the values discussed above (cathode particle radius of 1□m and Li$_{x=1}$CoO$_2$ density of 5.05 g/cm$^3$) we predict the normalized specific surface area of the cathode particles to be 0.6 m$^2$/g. This number is also consistent with the value reported in Ref. [101] of 0.6 m$^2$/g, which was obtained from BET measurement. Considering that 50% of the total surface is non-basal plane, the normalized active surface area for Li transport is 0.3 m$^2$/g. From this value we estimate the active cathode surface area of a typical CR-2032 cell to be 3000(cm$^2$/g) x 0.011 (g) = 33 cm$^2$. So, the active surface area to geometric surface ratio □ is 33 (cm$^2$)/1.6(cm$^2$) = 21. We will use this □ value in the fitting work for Ref. [49] and [50].

The fits in this section are clearly quite approximate, with errors introduced both from the approximate methods of extracting the published data, determining $J_{active}$ and □, and lack of rigorous Ohmic behavior of the contributions assigned to the coating. In general, as data for each coating thickness originates from a different sample and in some cases the coating thickness can be very small, extracting the coating resistance is highly uncertain. All studies are for Al$_2$O$_3$ ALD coated cathodes and therefore we assume the coating material is similar and at least nominally am-Al$_2$O$_3$. These values therefore give a useful range for qualitative guidance on the effective resistance provided by ALD am-Al$_2$O$_3$ coatings measured to date. The set of all resistivity values are summarized in Table AIII-1 and the details of fitting for each data set are given below.

**Table AIII-1.** Summary of all resistivity data from previous experimental measurements used in the linear fitting.

| References | Resistivity (charge) MΩm | Resistivity (discharge) MΩm |
|---|---|---|
| Cheng et al. Ref. [47] | 7.8±0.22 | 15.4±1.3 |
| Li et al. Ref. [49] | 913±243 | N/A |
| Riley et al. Ref. [50] | 55 | N/A |
| Woo et al. Ref. [51] | 375±24 | 708.3±24.1 |

**Fitting details for Ref. [47]**

The data from this reference was extracted from the dQ/dV vs. V plots at different coating thicknesses, and the current density is fixed at 0.2C rate. As discussed above, the $J_{active}$ corresponding to 1C rate is 0.046mA/cm$^2$. So the $J_{active}$ corresponding to 0.2C rate is 0.0092 mA/cm$^2$. The voltage values of the dQ/dV peaks are taken to be the measured voltage V in the linear fits. Due to the use of a constant current in the experiment we are forced to combine the $R_{system}S_{system}J_{active}$ term and $V_0$ term together and use $V_1 = V_0 + R_{system}S_{system}J_{active}$ as the y-axis intercept of the fit, as discussed previously.

**Table AIII-2** Potential V (applied voltage in the external circuit), current density $J_{active}$ and thickness $L$ extracted from Ref. [47] and the corresponding fitting results. Here we define the discharging current to be negative and charging current to be positive.

| Discharging process | | | | |
|---|---|---|---|---|
| Potential (V) | $J_{active}$ (mA/cm$^2$) | Thickness (nm) | Fitting result | |
| 3.885 | -0.0092 | 2.5 | $\rho_{coating}$ ± standard error σ (MΩm) | $V_I$ ± standard error σ (V) |
| 3.867 | -0.0092 | 12.5 | 15.4±1.3 | 3.88±0.008 |
| 3.831 | -0.0092 | 25 | coefficient of determination R$^2$ | 0.986 |
| 3.706 | -0.0092 | 125 | | |

| Charging process | | | | |
|---|---|---|---|---|
| Potential (V) | $J_{active}$ (mA/cm$^2$) | Thickness (nm) | Fitting result | |
| 3.971 | 0.0092 | 2.5 | $\rho_{coating}$ ± standard error σ (MΩm) | $V_I$ ± standard error σ (V) |
| 3.981 | 0.0092 | 12.5 | 7.8±0.22 | 3.97±0.0013 |
| 3.991 | 0.0092 | 25 | coefficient of determination R$^2$ | 0.998 |
| 4.061 | 0.0092 | 125 | | |

The coating overpotential ($|V - V_I|$) versus the $|L_{coating}J_{active}|$ is shown in Fig. AIII-1.

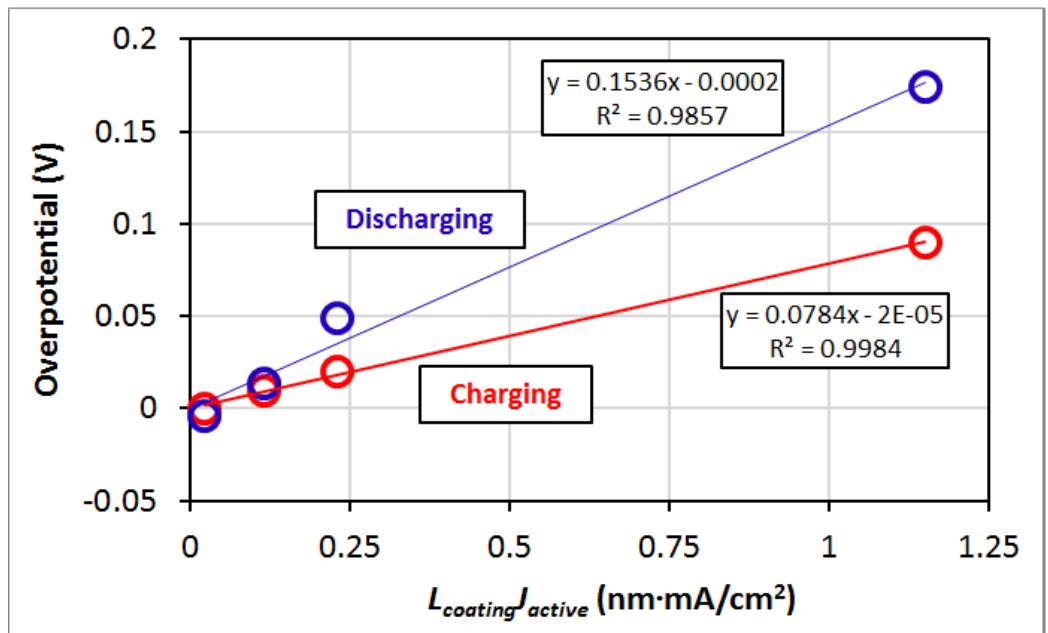

**Fig. AIII-1** Overpotential of the coating ($|V - V_I|$) vs. $|L_{coating}J_{active}|$. The slope corresponds to the coating resistivity.

Based on our fitting to the data from Ref. [47], we can see that the resistivity of the am-Al$_2$O$_3$ coating is 7.8±0.22 MΩm (1 MΩm = 10$^6$ Ωm) and 15.4±1.3 MΩm fitted from charging and discharging processes, respectively.

**Fitting details for Ref. [49]**

The data from this reference were extracted from the cyclic voltammogram plots at different coating thicknesses. The voltage values of the I-V curve peaks are taken to be the measured voltage $V$ in the linear fits. To calculate the $J_{active}$ corresponding to the I-V peaks, we take the total current values $I$ of the peaks directly from the original figures. As the author used CR-2032 type coin cell, we take the geometric surface area $A_{geom}$ = 1.6 cm$^2$ and the active surface area to geometric surface ratio $\Box$ = 21, based on our previous discussions. Based on these values, we can calculate $J_{active}$ and perform the Ohmic linear fitting.

**Table AIII-3** Potential $V$ (applied voltage in the external circuit), total current $I$, current density $J_{active}$ and thickness $L$ extracted from Ref. [49] and the corresponding fitting results.

| Discharging process | | | | | | |
|---|---|---|---|---|---|---|
| Potential (V) | total current $I$ (mA) | $J_{active}$ (mA/cm$^2$) | Thickness (nm) | Fitting result | | |
| 3.77 | -0.368 | -0.011 | 0.264 | $\rho_{coating}$ ± standard error σ (MΩm) | $R_{system}S_{system}$ ± standard error σ (MΩcm$^2$) | $V_0$ ±standard error σ (V) |
| 3.615 | -0.384 | -0.0114 | 0.66 | 2619±1224 | 0.078± 0.031 | 4.73±0.4 |
| 3.8 | -0.288 | -0.0086 | 1.32 | coefficient of determination R$^2$ | 0.88 | |
| 3.8 | -0.122 | -0.0036 | 6.6 | | | |

| Charging process | | | | | | |
|---|---|---|---|---|---|---|
| Potential (V) | total current $I$ (mA) | $J_{active}$ (mA/cm$^2$) | Thickness (nm) | Fitting result | | |
| 4.12 | 0.7 | 0.0208 | 0.264 | $\rho_{coating}$ ± standard error σ (MΩm) | $R_{system}S_{system}$ ± standard error σ (MΩcm$^2$) | $V_0$ ±standard error σ (V) |
| 4.25 | 0.82 | 0.0244 | 0.66 | 913±243 | 0.00077± 0.0046 | 4.06±0.13 |
| 4.29 | 0.667 | 0.0199 | 1.32 | coefficient of determination R$^2$ | 0.977 | |
| 4.45 | 0.21 | 0.0063 | 6.6 | | | |

We do not use the fitting results from the discharging data because the overpotential does not change as expected from our simple model. First, the overpotential does not change linearly with thickness in a robust manner. More significantly, if we compare the third or the fourth data point to the first data point, we can see that the potential $V$ actually increases when the coating thickness increases, which is contrary to our expectation that a thicker coating causes a larger overpotential, leading to a decrease in the measured voltage $V$. Therefore, we only use the charging process experimental data to fit the coating resistivity.

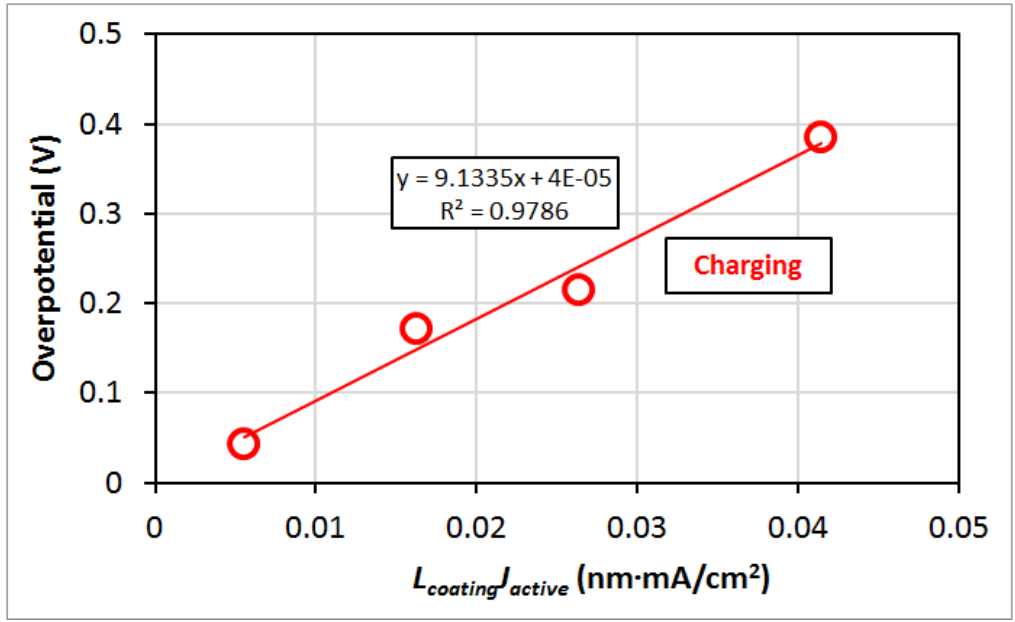

**Fig. AIII-2** Overpotential of the coating ($|V - V_0 - R_{system}S_{system}J_{active}|$) vs. $|L_{coating}J_{active}|$. The slope corresponds to the coating resistivity.

Therefore, based on the fitting work of Ref. [49], the resistivity of the am-$Al_2O_3$ coating is 913±243 MΩm for the charging process.

**Fitting details for Ref. [50]**

The data from this reference were extracted from the cyclic voltammogram plots at different coating thicknesses. The voltage values of the I-V curve peaks are taken to be the measured voltage $V$ in the linear fits. To calculate the $J_{active}$ corresponding to the I-V peaks, we read the total current values $I$ of the peaks directly from the original figures. As the author used CR-2032 type coin cell, we know the geometric surface area $A_{geo}$=1.6cm². We use the active surface area to geometric surface ratio $\square$ = 21 determined above. We can then calculate the $J_{active}$ values and perform the required linear fitting.

In Ref. [50], the discharging data have a similar problem with Ref. [49], as there are two data points whose voltage do not go down when the coating thickness increases, which is contrary to our expectation for the discharging process. Therefore, we only fit the experimental data of the charging process. There is one outlier point in the charging data and we exclude that point during the fitting.

**Table AIII-4** Potential $V$ (applied voltage in the external circuit), total current $I$, current density $J_{active}$ and thickness $L$ extracted from Ref. [50] and the corresponding fitting results.

| Charging process | | | | | | |
|---|---|---|---|---|---|---|
| Potential (V) | total current $I$ (mA) | $J_{active}$ (mA/cm²) | Thickness (nm) | Fitting result | | |
| 3.9 | 1.1 | 0.0327 | 0 | $\rho_{coating}$ (MΩm) | $R_{system}S_{system}$ (MΩcm²) | $V_0$ (V) |
| 3.91 | 0.88 | 0.026 | 0.8 | 55 | 0.00017 | 3.89 |
| 3.92 | 1.03 | 0.0307 | 1.2 | | | |

We do not have standard error and coefficient of determination $R^2$ in this case because we only have three data points from the Ref. [50] measurement, and in Eq. AIII-1 there are three parameters to fit. The am-$Al_2O_3$ resistivity based on the measurement of Ref. [50] is 55 MΩm.

**Fitting details for Ref. [51]**

The data from this reference were extracted from the voltage-capacity plots at different coating thicknesses. The point with the largest slope of $dQ/dV$ gives us the measured voltage $V$ in the linear fits. In this work the charging and discharging current is fixed at the geometric current density of $J_{geom}$= 0.045 mA/cm$^2$. The geometric surface area of the cell used in the experiment is $A_{geom}$=1.33 cm$^2$. The active $LiCoO_2$ cathode material in the cell is 3.77 mg. Based on the analysis above we take the specific surface area of $LiCoO_2$ cathode particles active for Li transport as 0.3 m$^2$/g. So the total active surface area is $A_{active}$=3000(cm$^2$/g) x 0.00377(g) =11.31cm$^2$. Thus we can calculate the $J_{active}$= $J_{geom}$x $A_{geom}/A_{active}$ = 0.045(mA/cm$^2$) x 1.33(cm$^2$) / 11.31(cm$^2$) = 0.0053 mA/cm$^2$. Due to the use of a constant current in the experiment we are forced to combine the $R_{system}S_{system}J_{active}$ term and $V_0$ term together and use $V_1 = V_0 + R_{system}S_{system}J_{active}$ as the y-axis intercept of the fit, as discussed previously.

**Table AIII-5** Potential $V$ (applied voltage in the external circuit), geometric current density $J_{geom}$, active current density $J_{active}$ and thickness $L$ extracted from Ref. [51] and the corresponding fitting results.

| Discharging process ||||||
|---|---|---|---|---|---|
| Potential (V) | $J_{geom}$ (mA/cm$^2$) | $J_{active}$ (mA/cm$^2$) | Thickness (nm) | Fitting result ||
| 3.86 | -0.045 | -0.0053 | 0 | $\rho_{coating}$± standard error σ (MΩm) | $V_1$ ±standard error σ (V) |
| 3.852 | -0.045 | -0.0053 | 0.23 | 708.3±24.1 | 3.86±0.0004 |
| 3.843 | -0.045 | -0.0053 | 0.46 | coefficient of determination $R^2$ | 0.999 |

| Charging process ||||||
|---|---|---|---|---|---|
| Potential (V) | $J_{geom}$ (mA/cm$^2$) | $J_{active}$ (mA/cm$^2$) | Thickness (nm) | Fitting result ||
| 3.961 | 0.045 | 0.0053 | 0 | $\rho_{coating}$± standard error σ (MΩm) | $V_1$ ±standard error σ (V) |
| 3.965 | 0.045 | 0.0053 | 0.23 | 375±24 | 3.96±0.0004 |
| 3.97 | 0.045 | 0.0053 | 0.46 | coefficient of determination $R^2$ | 0.996 |

So the resistivity of am-$Al_2O_3$ coating is 375±24 MΩm and 708.3±24.1 MΩm for the charging process and discharging process, respectively.

Table of Contents Figure (for ToC only):

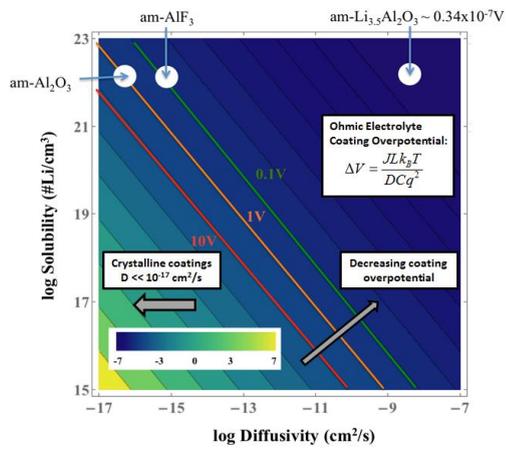